\documentclass[12pt]{article}
\usepackage{amssymb}

\def\hybrid{\topmargin 0pt      \oddsidemargin 0pt
        \headheight 0pt \headsep 0pt
        \voffset=-0.5cm
        \hoffset=-0.25in
        \textwidth 6.75in
        \textheight 9.5in       
        \marginparwidth 0.0in
        \parskip 5pt plus 1pt   \jot = 1.5ex}
\catcode`\@=11
\def\marginnote#1{}

\newcount\hour
\newcount\minute
\newtoks\amorpm
\hour=\time\divide\hour by60 \minute=\time{\multiply\hour by60
\global\advance\minute by-\hour}
\edef\standardtime{{\ifnum\hour<12 \global\amorpm={am}%
        \else\global\amorpm={pm}\advance\hour by-12 \fi
        \ifnum\hour=0 \hour=12 \fi
        \number\hour:\ifnum\minute<10 0\fi\number\minute\the\amorpm}}
\edef\militarytime{\number\hour:\ifnum\minute<10 0\fi\number\minute}

\def\draftlabel#1{{\@bsphack\if@filesw {\let\thepage\relax
   \xdef\@gtempa{\write\@auxout{\string
      \newlabel{#1}{{\@currentlabel}{\thepage}}}}}\@gtempa
   \if@nobreak \ifvmode\nobreak\fi\fi\fi\@esphack}
        \gdef\@eqnlabel{#1}}
\def\@eqnlabel{}
\def\@vacuum{}
\def\draftmarginnote#1{\marginpar{\raggedright\scriptsize\tt#1}}
\def\draftlabel#1{{\@bsphack\if@filesw {\let\thepage\relax
   \xdef\@gtempa{\write\@auxout{\string
      \newlabel{#1}{{\@currentlabel}{\thepage}}}}}\@gtempa
   \if@nobreak \ifvmode\nobreak\fi\fi\fi\@esphack}
        \gdef\@eqnlabel{#1}}
\def\@eqnlabel{}
\def\@vacuum{}
\def\draftmarginnote#1{\marginpar{\raggedright\scriptsize\tt#1}}

\def\draft{\oddsidemargin -.5truein
        \def\@oddfoot{\sl preliminary draft \hfil
        \rm\thepage\hfil\sl\today\quad\militarytime}
        \let\@evenfoot\@oddfoot \overfullrule 3pt
        \let\label=\draftlabel
        \let\marginnote=\draftmarginnote
   \def\@eqnnum{(\theequation)\rlap{\kern\marginparsep\tt\@eqnlabel}%
\global\let\@eqnlabel\@vacuum}  }

\def\numberbysection{\@addtoreset{equation}{section}
        \def\theequation{\thesection.\arabic{equation}}}

\def\underline#1{\relax\ifmmode\@@underline#1\else
        $\@@underline{\hbox{#1}}$\relax\fi}

\def\titlepage{\@restonecolfalse\if@twocolumn\@restonecoltrue\onecolumn
     \else \newpage \fi \thispagestyle{empty}\c@page\z@
        \def\thefootnote{\fnsymbol{footnote}} }

\def\endtitlepage{\if@restonecol\twocolumn \else  \fi
        \def\thefootnote{\arabic{footnote}}
        \setcounter{footnote}{0}}  

\numberbysection \hybrid

\newcounter{mo}

\newcommand{\tr}{{\rm tr}}
\newcommand{\ti}[1]{\tilde{#1}}

\newcommand{\mL}{{\mathcal L}}
\newcommand{\mM}{{\mathcal M}}
\newcommand{\mF}{{\mathcal F}}

\newcommand{\vf}{\varphi}
\newcommand{\al}{\alpha}
\newcommand{\be}{\beta}
\newcommand{\ga}{\gamma}
\newcommand{\om}{\omega}
\newcommand{\vth}{\vartheta}

\newcommand{\mC}{\mathbb C}
\newcommand{\mZ}{\mathbb Z}

\newcommand{\mS}{\mathcal S}

\newcommand{\ox}{\otimes}

\newcommand{\F}{\mathcal{F}}

\newtheorem{predl}{Proposition}[section]

\def\beq{\begin{equation}}
\def\eq{\end{equation}}
\def\p{\partial}

\begin{document}

\setcounter{page}{1}

\date{}
\date{}


\

\begin{center}
%
{\Large{On R-matrix-valued Lax pairs for Calogero-Moser models}}

\vspace{18mm}

{\large  {A. Grekov}\,\footnote{Steklov Mathematical Institute of
Russian Academy of Sciences, 8 Gubkina St., Moscow 119991, Russia;
e-mail: grekovandrew@mail.ru}
 \quad\quad\quad\quad
{A. Zotov}\,\footnote{Steklov Mathematical Institute of Russian
Academy of Sciences, 8 Gubkina St., Moscow 119991, Russia;
  e-mail:
zotov@mi.ras.ru}
 }
\end{center}

\vspace{1mm}


 \begin{abstract}
The article is devoted to the study of $R$-matrix-valued Lax pairs
for $N$-body (elliptic) Calogero-Moser models. Their matrix elements
are given by quantum ${\rm GL}_{\tilde N}$ $R$-matrices of
Baxter-Belavin type. For $\tilde N=1$ the widely known Krichever's
Lax pair with spectral parameter is reproduced. First, we construct
the $R$-matrix-valued Lax pairs for Calogero-Moser models associated
with classical root systems. For this purpose we study
generalizations of the D'Hoker-Phong Lax pairs. It appeared that in
the $R$-matrix-valued case the Lax pairs exist in special cases
only. The number of quantum spaces (on which $R$-matrices act) and
their dimension depend on the values of coupling constants. Some of
the obtained classical Lax pairs admit straightforward extension to
the quantum case. In the end we describe a relationship of the
$R$-matrix-valued Lax pairs to Hitchin systems defined on ${\rm
SL}_{N\tilde N}$ bundles with nontrivial characteristic classes over
elliptic curve. We show that the classical analogue of the
anisotropic spin exchange operator entering the $R$-matrix-valued
Lax equations is reproduced in these models.


 \end{abstract}


 \footnotesize \tableofcontents  \normalsize



\section{Introduction}
\setcounter{equation}{0}
In this paper we consider the Calogero-Moser models \cite{Ca} and
their generalizations of different types. The Hamiltonian of the
elliptic classical ${\rm sl}_N$ model
  \beq\label{a01}
  \begin{array}{c}
  \displaystyle{
H=\sum\limits_{i=1}^N\frac{p_i^2}{2}-\nu^2\sum\limits_{i>j}^N\wp(q_i-q_j)
 }
 \end{array}
 \eq
together with the canonical Poisson brackets
  \beq\label{a02}
  \begin{array}{c}
    \displaystyle{
\{p_i,q_j\}=\delta_{ij}\,,\quad \{p_i,p_j\}=\{q_i,q_j\}=0\,.
 }
 \end{array}
 \eq
 provides equations of motion for $N$-particle dynamics:
  \beq\label{a03}
  \begin{array}{c}
  \displaystyle{
 {\dot q}_i=p_i\,,\quad  {\ddot q}_i=\nu^2\sum\limits_{k: k\neq i}^N\wp'(q_{ik})\,.
 }
 \end{array}
 \eq
All variables and the coupling constant $\nu$ are assumed to be
complex numbers. Equations (\ref{a03}) can be written in the Lax
form. The Krichever's Lax pair with spectral parameter \cite{Krich1}
reads as follows\footnote{$\{E_{ij}\in{\rm Mat}(N),\ i,j=1...N\}$ --
is the standard basis in ${\rm Mat}(N)$:
$(E_{ij})_{kl}=\delta_{ik}\delta_{jl}$.}:
  \beq\label{a04}
  \begin{array}{c}
  \displaystyle{
L(z)=\sum\limits_{i,j=1}^N E_{ij}\,L_{ij}(z)\,,\quad
L_{ij}(z)=\delta_{ij}p_i+\nu(1-\delta_{ij})\phi(z,q_{ij})\,,\quad
q_{ij}=q_i-q_j\,,
 }
 \end{array}
 \eq
  \beq\label{a05}
  \begin{array}{c}
  \displaystyle{
M_{ij}(z)=\nu d_i\delta_{ij} +\nu(1-\delta_{ij})f(z,q_{ij})\,,\quad
d_i=\sum\limits_{k: k\neq i}^N E_2(q_{ik})=-\sum\limits_{k: k\neq
i}^N f(0,q_{ik})\,,
 }
 \end{array}
 \eq
i.e. the Lax equations
  \beq\label{a06}
  \begin{array}{c}
  \displaystyle{
{\dot L}(z)\equiv\{H,L(z)\}=[L(z),M(z)]
 }
 \end{array}
 \eq
are equivalent to (\ref{a03}) identically in $z$. The definitions
and properties of elliptic functions entering
(\ref{a01})-(\ref{a05}) are given in the Appendix. The proof is
based on the identity (\ref{a966}) written as
  \beq\label{a15}
  \begin{array}{c}
  \displaystyle{
\phi(z,q_{ab})f(z,q_{bc})-f(z,q_{ab})\phi(z,q_{bc})=
\phi(z,q_{ac})(f(0,q_{bc})-f(0,q_{ab}))\,.
 }
 \end{array}
 \eq
and
  \beq\label{a151}
  \begin{array}{c}
  \displaystyle{
\phi(z,q_{ab})f(z,q_{ba})-f(z,q_{ab})\phi(z,q_{ba})=\wp'(q_{ab})\,.
 }
 \end{array}
 \eq
These are particular cases of the genus one Fay identity
(\ref{a909})
  \beq\label{a14}
  \begin{array}{c}
  \displaystyle{
\phi(z,q_{ab})\phi(w,q_{bc})=\phi(w,q_{ac})\phi(z-w,q_{ab})+\phi(w-z,q_{bc})\phi(z,q_{ac})\,.
 }
 \end{array}
 \eq
The model (\ref{a01})-(\ref{a03}) is included into a wide class of
Calogero-Moser models associated with root systems
\cite{OP}. The corresponding Lax pairs with spectral parameter were
found in \cite{DP,BCS}. In particular, for the ${\rm BC}_N$ root
system described by the Hamiltonian
 \begin{equation}
 \label{a10}
 H = \frac{1}{2} \sum_{a=1}^{N} p_a^2 - \nu^2
\sum_{a<b}^{N} (\wp(q_a-q_b) + \wp(q_a+q_b)) - \mu^2\sum_{a=1}^{N}
\wp(2q_a) - g^2 \sum_{a=1}^{N} \wp(q_a)
 \end{equation}
there exists the Lax pair with spectral parameter of size
$(2N+1)\times(2N+1)$ if (as in \cite{OP})
 \begin{equation}
 \label{a102}
  g(g^2 - 2\nu^2 + \nu \mu) = 0\,.
 \end{equation}
Let us remark that the Lax pairs of size $3N\times 3N$ \cite{Inoz89}
or $2N\times 2N$ \cite{Feher2} corresponding to the general case
(all constants are arbitrary) are not considered in this paper.

The Lax pair (\ref{a04})-(\ref{a05}) of the ${\rm sl}_N$ model
(\ref{a01})-(\ref{a03}) has the following generalization \cite{LOZ}
(the $R$-matrix-valued Lax pair):\footnote{Equations of motion
following from (\ref{a07})-(\ref{a08}) contain the coupling constant
${\ti N}\nu$ instead of $\nu$ in (\ref{a01}), (\ref{a03}). }
  \beq\label{a07}
  \begin{array}{c}
    \displaystyle{
{\mL}(z)=\sum\limits_{i,j=1}^N E_{ij}\otimes \mL_{ij}(z)\,,
 \quad\quad
\mL_{ij}(z)= 1_{\ti N}^{\otimes N}\, \delta_{ij}p_i
+\nu(1-\delta_{ij})R^{\,z}_{ij}(q_{ij})
 }
 \end{array}
 \eq
  \beq\label{a08}
  \begin{array}{c}
  \displaystyle{
\mM_{ij}(z)=\nu d_i\delta_{ij}
+\nu(1-\delta_{ij})F^{\,z}_{ij}(q_{ij})+\nu\delta_{ij}\,\mF^{\,0}\,,\quad
d_i=-\sum\limits_{k: k\neq i}^N F^{\,0}_{ik}(q_{ik})\,,
 }
 \end{array}
 \eq
where $F^{\,z}_{ij}(q)=\p_q R^{\,z}_{ij}(q)$,
$F^{\,0}_{ij}(q)=F^{\,z}_{ij}(q)|_{z=0}=F^{\,0}_{ji}(-q)$
(\ref{a1007}) and
  \beq\label{a09}
  \begin{array}{c}
  \displaystyle{
\mF^{\,0}
 =\sum\limits_{k>m}^N F^{\,0}_{km}(q_{km})
 =\frac{1}{2}\sum\limits_{k,m=1}^N F^{\,0}_{km}(q_{km})\,.
 }
 \end{array}
 \eq
It has block-matrix structure\footnote{The operator-valued Lax pairs
with a similar structure are known
\cite{Inoz0,HW,HW2,Bernard,Inoz1}. We discuss it below.}. The blocks
are enumerated by $i,j=1...N$ as matrix elements in (\ref{a04}).
Each block of $\mL(z)$ is some ${\rm GL}(\ti N)$-valued $R$-matrix
in fundamental representation, acting on the $N$-th tensor power of
$\ti N$-dimensional vector space ${\mathcal H}=(\mC^{\ti
N})^{\otimes N}$. So the size of each block is ${\rm dim}{\mathcal
H}\times{\rm dim}{\mathcal H}$, and ${\rm dim}{\mathcal H}={\ti
N}^N$, i.e. $\mL(z)\in{\rm Mat}_N\otimes {\rm Mat}_{\ti N}^{\otimes
N}$. We will refer to ${\rm Mat}_N$ component as auxiliary space,
and to ${\rm Mat}_{\ti N}^{\otimes N}\cong{\mathcal H}^{\otimes 2}$
-- as "quantum" space since ${\mathcal H}$ is the Hilbert space of
${\rm GL}(\ti N)$ spin chain (in fundamental representation) on $N$
sites.

An $R$-matrix $R_{ij}$ acts trivially in all tensor components
except $i,j$. It is normalized in a way that for $\ti N=1$ it is
reduced to the Kronecker function $\phi(z,q_{ij})$ (\ref{a907})
\cite{Weil}. For instance, in one of the simplest examples $R_{ij}$
is the Yang's $R$-matrix \cite{Yang}:
  \beq\label{a13}
  \begin{array}{c}
  \displaystyle{
  R^{\eta}_{12}(q)=\frac{1\otimes
  1}{\eta}+\frac{{\ti N}P_{12}}{q}\,,
 }
 \end{array}
 \eq
where $P_{12}$ is the permutation operator (\ref{a9051}). In general
(and as a default) case $R_{ij}$ is the Baxter-Belavin
\cite{Baxter,Belavin} elliptic $R$-matrix (\ref{a1004}). The
properties of this $R$-matrix are very similar to those of the
function $\phi(z,q)$. The key equation for $R_{ij}$ (which is needed
for existence of the $R$-matrix-valued Lax pair) is the associative
Yang-Baxter equation \cite{FK}
  \beq\label{a16}
  \begin{array}{c}
  \displaystyle{
 R^z_{ab}
 R^{w}_{bc}=R^{w}_{ac}R_{ab}^{z-w}+R^{w-z}_{bc}R^z_{ac}\,,\
 \ R^z_{ab}=R^z_{ab}(q_a\!-\!q_b)\,.
 }
 \end{array}
 \eq
It is a matrix generalization of the Fay identity (\ref{a14}), and
it is fulfilled by the Baxter-Belavin $R$-matrix \cite{Pol}. The
degeneration of (\ref{a16}) similar to (\ref{a15}) is of the form:
  \beq\label{a17}
  \begin{array}{c}
  \displaystyle{
R^z_{ab} F^z_{bc} - F^z_{ab} R^z_{bc} = F^0_{bc} R^z_{ac} - R^z_{ac}
F^0_{ab}\,.
 }
 \end{array}
 \eq
It underlies the Lax equations for the Lax pair
(\ref{a07})-(\ref{a08}).
The last term $\mF^{\,0}$ in (\ref{a08}) is not needed in
(\ref{a05}) since for $\ti N=1$ it is proportional to the identity
$N\times N$ matrix. But it is important for $\ti N>1$ since it
changes the order of $R$ and $F^0$ in the r.h.s. of (\ref{a17}).
Namely,
  \beq\label{a12}
  \begin{array}{c}
  \displaystyle{
 [R^z_{ac},\mF^{\,0}]+\sum\limits_{b\neq a,c}R^z_{ab} F^z_{bc} -
F^z_{ab} R^z_{bc} = \sum\limits_{b\neq c} R^z_{ac} F^0_{bc} -
 \sum\limits_{b\neq a} F^0_{ab} R^z_{ac}\,,\quad  \forall\ a\neq c.
 }
 \end{array}
 \eq
 This identity provides cancellation of non-diagonal blocks in the
 Lax equations.
%
%
%
%
%
See \cite{Pol,Pol2,LOZ15,LOZ16,Unitary} for different properties and
applications of $R$-matrices of the described type. Here we need two
more important properties. These are the unitarity
  \beq\label{a18}
  \begin{array}{c}
  \displaystyle{
 R^z_{12}(q_{12}) R^{z}_{21}(q_{21})=1\otimes 1\,{\ti N}^2(\wp({\ti N}z)-\wp(q_{12}))
 }
 \end{array}
 \eq
and the skew-symmetry
 \beq\label{a19}
 \begin{array}{c}
  \displaystyle{
 R^z_{ab}(q)=-R^{-z}_{ba}(-q)\,.
 }
 \end{array}
 \eq
On the one hand these properties are needed for the Lax equations
since they lead to
 \beq\label{a191}
 \begin{array}{c}
  \displaystyle{
 F^0_{ab}(q)=F^{0}_{ba}(-q)
 }
 \end{array}
 \eq
 and to the analogue of (\ref{a151}) (obtained by differentiating the identity (\ref{a18}))
 \beq\label{a192}
 \begin{array}{c}
  \displaystyle{
 R^z_{ab} F^z_{ba} - F^z_{ab} R^z_{ba}={\ti N}^2\wp'(q_{ab})\,,
 }
 \end{array}
 \eq
which provides equations of motion in each diagonal block in the Lax
equations.

 On the other hand, together with (\ref{a18}) and (\ref{a19}) the associative
 Yang-Baxter equation leads to the quantum Yang-Baxter
 equation
 $R^\eta_{ab}R^\eta_{ac}R^\eta_{bc}=R^\eta_{bc}R^\eta_{ac}R^\eta_{ab}$.
In this respect we deal with the quantum $R$-matrices satisfying
(\ref{a16}), (\ref{a18}), (\ref{a19}), and the Planck constant of
$R$-matrix plays the role of the spectral parameter for the  Lax
pair (\ref{a07})-(\ref{a08}). In trigonometric case the $R$-matrices
satisfying the requirements include the standard ${\rm GL}(\ti N)$
XXZ $R$-matrix \cite{KSkl} and its deformation \cite{Chered,AHZ}
(${\rm GL}(\ti N)$ extension of the 7-vertex $R$-matrix). In the
rational case the set of the $R$-matrices includes the Yang's one
(\ref{a13}) and its deformations \cite{Chered,LOZ5} (${\rm GL}(\ti
N)$ extension of the 11-vertex $R$-matrix).


{\bf The aim of the paper} is to clarify the origin of the
$R$-matrix-valued Lax pairs and examine some known constructions,
which work for the ordinary Lax pairs.

First, we study extensions of (\ref{a07})-(\ref{a08}) to other root
systems. More precisely, we propose $R$-matrix-valued extensions of
the D'Hoker-Phong Lax pairs for (untwisted) Calogero-Moser models
associated with classical root systems and ${\rm BC}_N$ (\ref{a10}).
The auxiliary space in these cases is given by ${\rm Mat}_{2N}$ or
${\rm Mat}_{2N+1}$ because such root systems are obtained from ${\rm
sl}_{2N}$ or ${\rm sl}_{2N+1}$ cases by discrete reduction. There
are two natural possibilities for arranging tensor components of the
quantum spaces. The first one is to keep $2N+1$ (or $2N$) components
of the quantum spaces in the reduced root system. The second -- is
to leave only $N$ (or $N+1$) components. We study both cases.

Next, we proceed to quantum Calogero-Moser models
\cite{Ca,OP1,Chered0}. To some extent they are described by quantum
analogue of the Lax equations (\ref{a06}) \cite{Ujino,BMS}:
  \beq\label{a20}
  \begin{array}{c}
  \displaystyle{
[\hat{H},\hat{L}(z)]=\hbar\,[\hat{L}(z),M(z)]\,,
 }
 \end{array}
 \eq
 where $\hat{H}$ is the quantum Hamiltonian (it is scalar in the auxiliary space), $\hat{L}(z)$ is the
 quantum Lax matrix and $\hbar$ is the Planck constant. The operators $\hat{H}$ and $\hat{L}(z)$
 are obtained from the classical
 (\ref{a01}) and (\ref{a04}) by replacing momenta $p_i$ with
 $\hbar\,\p_{q_i}$, and the coupling constant in the Hamiltonian acquires the quantum correction.
 We verify if the obtained $R$-matrix-valued Lax pairs are
 generalized to quantum case in a similar way. It appears that (besides the ${\rm sl}_N$ case) only
 models associated with
 ${\rm SO}$ type root systems are generalized.
 %
As a result we show\footnote{Some more details are given in the
Conclusion.}
 \begin{predl}
The D'Hoker-Phong Lax pairs for (untwisted) classical Calogero-Moser
models associated with classical root systems and ${\rm BC}_N$ admit
$R$-matrix-valued extensions with additional constraints:

-- for the coupling constants in ${\rm C}_N$ and ${\rm BC}_N$ cases;

-- for the size of $R$-matrix ($\ti N=2$) in ${\rm B}_N$ and ${\rm
D}_N$ cases.

\noindent The latter cases are directly generalized to quantum Lax
equations, while the ${\rm C}_N$ and ${\rm BC}_N$ cases are not. The
${\rm A}_N$ Lax pair is generalized to the quantum case
straightforwardly without any restrictions.
 \end{predl}

The Calogero-Moser models \cite{Ca,OP1} possess also spin
generalizations \cite{GH}. Its Lax description is known at classical
\cite{BAB} and quantum levels \cite{Inoz0,HW,HW2,Bernard,Inoz1}. It
is important to note that for the quantum Calogero-Moser models
 with spin  the quantum Lax pairs have the same operator-valued
(tensor) structure as in (\ref{a07})-(\ref{a08}). The term analogues
to $\mF^0$ (\ref{a09}) is treated as a part of the quantum
Hamiltonian, describing interaction of spins. We explain (in Section
\ref{quant}) how the $R$-matrix-valued Lax pairs generalize (and
reproduce) the previously known results.

Finally, we discuss an origin of the $R$-matrix-valued Lax pairs
(for  ${\rm sl}_N$ case (\ref{a07})-(\ref{a08}) with ${\rm GL}_{\ti
N}$ $R$-matrices) by relating them to Hitchin systems on ${\rm
SL}(N\ti N)$-bundles over elliptic curve. Originally systems of this
type
were derived by A. Polychronakos from matrix models \cite{Polych}
and later were described as Hitchin systems with nontrivial
characteristic classes \cite{LZ,LOSZ}. It is also known as the model
of interacting tops since it is Hamiltonian (or equations of motion)
are treated as interaction of $N$ ${\rm SL}(\ti N)$-valued elliptic
tops.

The relation between the $R$-matrix-valued Lax pairs and the
interacting tops comes from rewriting
the Lax equation for (\ref{a07})-(\ref{a08}) in the form
  \beq\label{a2020}
  \begin{array}{c}
  \displaystyle{
\{H,\mL\}+[\nu\mF^0,\mL(z)]=[\mL(z),\bar{\mM}(z)]\,,
 }
 \end{array}
 \eq
 where in contrast to (\ref{a08}) $\bar{\mM}$ does not include the $\mF^0$ term (\ref{a09}).
 In this respect the $R$-matrix-valued Lax pair is "half-quantum":
 the spin variables are quantized in the fundamental representation,
 while the positions and momenta of particles remain classical. The $\mF^0$ term
 in this treatment is the (anisotropic) spin exchange operator.
We will show that the classical analogue for such spin exchange
operator appear in the above mentioned Hitchin systems.
Alternatively, the result is formulated as follows.
 \begin{predl}\label{2222}
The quantum Hamiltonian $\hat{H}^{\rm tops}$ of the model of $N$
interacting ${\rm SL}(\ti N)$ elliptic tops (with spin variables
being quantized in the fundamental representation) coincides with
 the sum of the quantum Calogero-Moser Hamiltonian (\ref{a20}) and
 $\mF^0$-term (\ref{a09})
  \beq\label{a2222}
  \begin{array}{c}
  \displaystyle{
\hat{H}^{\rm tops}=\hat{H}^{\rm CM}+\hbar\nu\mF^0+1_{\ti N}^{\otimes
N}\hbox{const}
 }
 \end{array}
 \eq
 up to a constant proportional to identity matrix in ${\rm End}(\mathcal H)$ and redefinition of the coupling constants.
 \end{predl}
We prove this Proposition in the end of Section 4.











\section{Classical root systems}\label{root}
\setcounter{equation}{0}
 In \cite{DP} the following Lax pair was found for the model
 (\ref{a10}):
 \begin{equation}\label{a31}
L =\left(
    \begin{array}{ccc}
    P+A & B_1 & C_1     \\
    B_2 & -P+A^T & C_2    \\
    C_2^T & C_1^T & 0
        \end{array}
    \right)
 \quad\quad\quad
 M =\left(
    \begin{array}{ccc}
    A'+d & B_1' & C_1'    \\
    B_2' & A'^T+d & C_2' \\
    C_2'^T & C_1'^T & d_0 \\
        \end{array}
     \right)
 \end{equation}
where $A,B,P,D\in{\rm Mat}_N$, $C_1,C_2$ are columns of length $N$
and
 \beq\label{a321}
  \begin{array}{c}
  \displaystyle{P_{ab} = \delta_{ab}p_a}\,,\quad\quad\quad
  \displaystyle{A_{ab} = \nu (1-\delta_{ab})\phi(z,q_a-q_b)}\,,\\ \ \\
  \displaystyle{(B_1)_{ab} = \nu (1-\delta_{ab})\phi(z,q_a+q_b) + \mu \delta_{ab} \phi(z,2q_a)}\,,\\ \ \\
  \displaystyle{(B_2)_{ab} = \nu (1-\delta_{ab})\phi(z,-q_a-q_b) + \mu \delta_{ab} \phi(z,-2q_a)}\,,\\ \ \\
  \end{array}
 \eq
 \beq\label{a322}
  \begin{array}{c}
  \displaystyle{(C_1)_{a} = g\phi(z,q_a)}\,,\quad\quad\quad
  \displaystyle{(C_2)_{a} = g\phi(z,-q_a)}\,,\\ \ \\
  \displaystyle{d_{ab}=\delta_{ab}d_a\,,\quad
  d_{a} = \frac{g^2}{\nu} \wp(q_a) +\mu \wp(2q_a) +\nu \sum_{b \neq a} (\wp(q_a-q_b)+\wp(q_a+q_b))}\,,\\
  \displaystyle{d_0 = 2\nu \sum_c \wp(q_c)}\,.
  \end{array}
 \eq
The superscript $T$  stands for transposition, and the prime means
the derivative with respect to the second argument of functions
$\phi(z,q)$, i.e. $\phi(z,q)$ are replaced by $f(z,q)$ as in
(\ref{a05}).

The Lax pair (\ref{a31}), (\ref{a321}), (\ref{a322}) satisfies the
Lax equations (\ref{a06}) with the Hamiltonian (\ref{a10}) if the
additional constraint (\ref{a102}) for the coupling constants $\nu$,
$\mu$ and $g$ holds true. In the following particular cases the
classical root systems arise:

-- ${\rm B}_N$ (${\rm so}_{2N+1}$): $\mu=0$, $g^2=2\nu^2$;

-- ${\rm C}_N$ (${\rm sp}_{2N}$): $g=0$;

-- ${\rm D}_N$ (${\rm so}_{2N}$): $\mu=0$, $g=0$.

\noindent In order to deal with $2N\times 2N$ matrices in ${\rm
C}_N$ and ${\rm D}_N$ cases one may subtract $d_0 1_{2N+1}$ from the
$M$-matrix.

Our purpose is to generalize (\ref{a31}), (\ref{a321}), (\ref{a322})
to $R$-matrix-valued Lax pair of (\ref{a07}), (\ref{a08}) type. The
problem is that $R_{ij}^z(q)$ carries the indices, which enumerate
tensor components in quantum space. Hence we should define a number
of these components and arrange them. There are two natural
possibilities to do it\footnote{There is also a kind of intermediate
case in the so-called universal Lax pairs \cite{Inoz1}. The number
of quantum spaces (sites) is equal to the Lax matrix size, and the
$R$-matrix is proportional to the permutation operator defined by
Weyl group elements of the corresponding root system. For example,
in ${\rm D}_N$ case the permutation of $i$-th and $j$-th sites
($1\leq i,j\leq N$) includes also permutation of $i+N$-th and
$j+N$-th sites. Elliptic $R$-matrix exists for ${\rm GL}$ case only.
For this reason we do not know generalization for these operators to
(elliptic) $R$-matrix satisfying the necessary conditions.}:

\begin{itemize}
 \item the first one is to keep the number of components to be equal to
the Lax matrix size, i.e. to $2N+1$ (or, to $2N$ if $g=0$).

 \item the second possibility is to leave only half of this set (coming
from ${\rm sl}_{2N+1}$) likewise it is performed for spin chains
with boundaries \cite{Skl2,BPS}. Put it differently, the number of
"spin sites" in quantum space is equal to the rank of the root
systems.
\end{itemize}
%

The first possibility leads to  $\mu=\nu$, and we are left with
${\rm BC}_N$ and ${\rm C}_N$ cases. They are described by
straightforward reductions from ${\rm sl}_{2N+1}$ and ${\rm
sl}_{2N}$ respectively\footnote{For example, the reduction from
${\rm sl}_{2N}$
 with positions of particles $u_i$ $i=1...2N$ to ${\rm C}_N$ is achieved
by identifying $u_i=q_i$ and $u_{i+N}=-q_i$, $i=1...N$  (and the
same for momenta). }, so in this case $g=\nu$ as well.  At the same
time the case $g=-\nu$ for ${\rm BC}_N$ root system also works. See
the explanation below (\ref{a387}).

The second possibility implies $\mu=0$ by the following reason. The
diagonal elements in the  blocks $B_1$ and $B_2$ (standing behind
$E_{a,N+a}$ and $E_{N+a,a}$, $a=1,...N$) should have elements with
$R$-matrices restricted to a single quantum space, i.e.
$R_{aa}^z(\pm2q_a)$. But such terms can not be involved into ansatz
based on (\ref{a16}), which is identity in ${\rm Mat}_{\ti
N}^{\otimes 3}$. Thus we are left with ${\rm B}_N$ and ${\rm D}_N$
cases. Moreover, we will see that only $\ti N=2$ is possible and $g
= \pm\sqrt{2}\nu$.

\subsection{${\rm C}_N$ case}
In this case $g=0$ and $\mu = \nu$, the auxiliary space is ${\rm
Mat}_{2N}$, the quantum space is ${\rm Mat}_{\ti N}^{\otimes 2N}$,
and $\ti N$ is arbitrary. The Lax pair:
 \begin{equation}\label{SO(2N)}
\mathcal{L} =
     \left(\begin{array}{cc}
    P+A_1 & B_1           \\
    B_2 & -P+A_2
         \end{array}
         \right)
\quad\quad
 \mathcal{M} =
      \left(\begin{array}{cc}
    A_1'+D_1+\mathcal{F} & B_1'          \\
    B_2' & A_2'+D_2+\mathcal{F}
         \end{array}\right)
 \end{equation}
where
  \beq\label{a33}
  \begin{array}{c}
  P = \sum\limits_a p_a E_{aa} \ox 1_{\ti N}^{\otimes 2N}\,,\\
A_1 = \nu \sum\limits_{a,b} E_{ab} \ox
R^{z}_{ab}(q_a-q_b)\,,\quad\quad
A_2 = \nu \sum\limits_{a,b} E_{ab} \ox R^{z}_{a+N,b+N}(-q_a+q_b)\,,\\
\quad
 B_1 = \nu \sum\limits_{a,b} E_{ab} \ox
R^{z}_{a,b+N}(q_a+q_b)\,,\quad\quad
B_2 = \nu \sum\limits_{a,b} E_{ab} \ox R^{z}_{a+N,b}(-q_a-q_b)\,,\\
A_1' = \nu \sum\limits_{a,b} E_{ab} \ox
F^{z}_{ab}(q_a-q_b)\,,\quad\quad
A_2' = \nu \sum\limits_{a,b} E_{ab} \ox F^{z}_{a+N,b+N}(-q_a+q_b)\,,\\
B_1' = \nu \sum\limits_{a,b} E_{ab} \ox
F^{z}_{a,b+N}(q_a+q_b)\,,\quad\quad B_2' = \nu \sum\limits_{a,b}
E_{ab} \ox F^{z}_{a+N,b}(-q_a-q_b)\,,
 \end{array}
 \eq
  \beq\label{a331}
  \begin{array}{c}
  D_1 = \nu \sum\limits_{a} E_{aa} \ox
  d_a\,,\quad\quad
d_a = -\sum\limits_{c:\,c\neq a} F^0_{ac}(q_a-q_c)-\sum\limits_{c} F^0_{a,c+N}(q_a+q_c)\,,\\
D_2 = \nu \sum\limits_{a} E_{aa} \ox d_{a+N}\,,\quad\quad d_{a+N} =
-\sum\limits_{c} F^0_{a+N,c}(q_a+q_c)-\sum\limits_{c:\, c \neq a}
F^0_{a+N,c+N}(q_a-q_c)
 \end{array}
 \eq
and $\mathcal{F}=\nu 1_N \ox \F^0$ with
  \beq\label{a34}
  \begin{array}{c}
\F^0 = \frac{1}{2}\sum\limits_{c \neq d}
(F^0_{cd}(q_c-q_b)+F^0_{c+N,d+N}(q_c-q_d))+\frac{1}{2}\sum\limits_{c,d}
(F^0_{c,d+N}(q_c+q_d)+F^0_{c+N,d}(q_c+q_d))\,.
 \end{array}
 \eq

\subsection{${\rm BC}_N$ case} Here $\pm g=\mu = \nu$, the auxiliary
space is ${\rm Mat}_{2N+1}$, the quantum space is ${\rm Mat}_{\ti
N}^{\otimes(2N+1)}$, and $\ti N$ is arbitrary.
 \begin{equation}
\label{BC(N)} \mathcal{L} =
     \left(\begin{array}{ccc}
    P+A_1 & B_1 & C_1     \\
    B_2 & -P+A_2 & C_2    \\
    C_2^T & C_1^T & 0
         \end{array}\right)
 \quad\quad
\mathcal{M} =
     \left(\begin{array}{ccc}
    A_1'+D_1+\mathcal{F} & B_1' & C_1'    \\
    B_2' & A_2'+D_2+\mathcal{F} & C_2' \\
    C_2'^T & C_1'^T & D_3+\mathcal{F}
         \end{array}\right)
 \end{equation}
where the blocks $A$, $B$ and $P$ are the same as in ${\rm C}_N$
case and
  \beq\label{a35}
  \begin{array}{c}
\left(C_1\right)_a = \pm \nu R^{z}_{a,2N+1}(q_a)\,,\quad\quad
\left(C_2\right)_a = \pm \nu  R^{z}_{a+N,2N+1}(-q_a)\,,\\ \ \\
\left(C_1^T\right)_a = \pm \nu R^{z}_{2N+1,a+N}(q_a)\,,\quad\quad
\left(C_2^T\right)_a = \pm \nu R^{z}_{2N+1,a}(-q_a)\,,\\ \ \\
\left(C_1'\right)_a = \pm \nu F^{z}_{a,2N+1}(q_a)\,,\quad\quad
\left(C_2'\right)_a = \pm \nu  F^{z}_{a+N,2N+1}(-q_a)\,,\\ \ \\
\left(C_1'^T\right)_a = \pm \nu F^{z}_{2N+1,a+N}(q_a)\,,\quad\quad
\left(C_2'^T\right)_a = \pm \nu  F^{z}_{2N+1,a}(-q_a)\,,\\ \ \\
 D_3 = \nu d_{2N+1}\,,\quad\quad
  d_{2N+1} =
-\sum\limits_{c} F^0_{c,2N+1}(q_c)-\sum\limits_{c}
F^0_{c+N,2N+1}(q_c)\,.
 \end{array}
 \eq
The  blocks $D_1 = \nu \sum\limits_{a} E_{aa} \ox d_a$, $D_2 = \nu
\sum\limits_{a} E_{aa} \ox d_{a+N}$ and $\mathcal{F}=\nu 1_N \ox
\F^0$ are given by:
  \beq\label{a36}
  \begin{array}{c}
 d_a = -\sum\limits_{c:\,c\neq a} F^0_{ac}(q_a-q_c)-\sum\limits_{c} F^0_{a,c+N}(q_a+q_c) - F^0_{a,2N+1}(q_a)\,,\\ \ \\
 d_{a+N} = -\sum\limits_{c} F^0_{a+N,c}(q_a+q_c)-\sum\limits_{c:\, c
\neq a} F^0_{a+N,c+N}(q_a-q_c) - F^0_{a+N,2N+1}(q_a)
 \end{array}
 \eq
and
  \beq\label{a37}
  \begin{array}{c}
  \displaystyle{
 \F^0 = \frac{1}{2}\sum\limits_{c \neq d}
 (F^0_{cd}(q_c-q_b)+F^0_{c+N,d+N}(q_c-q_d))+
 }
 \\
 \displaystyle{
 +\frac{1}{2}\sum\limits_{c,d}
 (F^0_{c,d+N}(q_c+q_d)+F^0_{c+N,d}(q_c+q_d))
 +\sum\limits_{c} F^0_{c,2N+1}(q_c)+\sum\limits_{c}
 F^0_{c+N,2N+1}(q_c) \,.}
 \end{array}
 \eq
For shortness we can also write (\ref{a36})-(\ref{a37}) as
  \beq\label{a38}
  \begin{array}{c}
d_a^{BC(N)} = d_a^{Sp(2N)} - F^0_{a,2N+1}(q_a)\,,\quad
d_{a+N}^{BC(N)} = d_{a+N}^{Sp(2N)} - F^0_{a+N,2N+1}(q_a)\,,\\ \ \\
\F^0_{BC(N)} = \F^0_{Sp(2N)} - d_{2N+1}\,.
 \end{array}
 \eq

Let us also comment on the necessity of $\pm g =\mu =\nu$. Consider,
for example, the block \{13\} of the Lax equation: \
  \beq\label{a381}
  \begin{array}{c}
 \displaystyle{ [\mL,\mM]^{13} = PC_1'+ A C_1' + B_1 C_2' + C_1 D_3 + [C_1, \F] - A'
C_1 - D_1 C_1 - B_1' C_2\,.
 }
 \end{array}
 \eq
The term $PC_1'$ gives the equation of motion. Next,
  \beq\label{a382}
  \begin{array}{c}
   \displaystyle{
(A C_1' - A' C_1)_a = \nu g \sum_{b} R^{z}_{ab}(q_a-q_b)
F^{z}_{b,2N+1}(q_b)- F^{z}_{ab}(q_a-q_b) R^{z}_{b,2N+1}(q_b) =}
 \\ \ \\
 \displaystyle{
 =\nu g \sum_{b} F^{0}_{ab}(q_a-q_b)R^{z}_{b,2N+1}(q_b) -
R^{z}_{b,2N+1}(q_b) F^0_{ab}(q_a-q_b)\,.
 }
 \end{array}
 \eq
 The rest of the terms (after applying the unitarity condition) yield:
  \beq\label{a383}
  \begin{array}{c}
[\mL,\mM]^{13}_a = \hbox{eq. of motion} +\\ \ \\+ \nu g \sum_{b: \,
b \neq a}
 [F^0_{b,2N+1}(q_b)R^{z}_{a,2N+1}(q_a) -
 R^{z}_{a,2N+1}(q_a)F^{0}_{ab}(q_a-q_b)]  +\\ \ \\+
 \nu g \sum_{b: \, b \neq a} [F^0_{b,2N+1}(q_b)R^{z}_{a,2N+1}(q_a)-
  R^{z}_{a,2N+1}(q_a)F^{0}_{a,b+N}(q_a+q_b)] +\\ \ \\+ \mu g (F^0_{a,2N+1}(q_a)R^{z}_{a,2N+1}(q_a)-
   R^{z}_{a,2N+1}(q_a)F^{0}_{a,b+N}(2q_a)) +\\ \ \\+
 \nu g R^{z}_{a,2N+1}(q_a) d_{2N+1} - g D_{1a}R^{z}_{a,2N+1}(q_a) +\nu g
 [R^{z}_{a,2N+1}(q_a),\F^0]\,.
 \end{array}
 \eq
Here $D_1$ and $\F^0$ have the following form:
  \beq\label{a384}
  \begin{array}{c}
D_{1a} = -\frac{g^2}{\nu} F^0_{a,2N+1} - \mu F^0_{a,a+N} - \nu
\sum_{b: \, b \neq a} F^0_{ab}(q_a-q_b) + F^0_{a,b+N}(q_a+q_b)\,,
 \end{array}
 \eq
  \beq\label{a385}
  \begin{array}{c}
    \displaystyle{
\F^0 = \frac{1}{2}\sum\limits_{c \neq d}
 \Big(F^0_{cd}(q_c-q_b)+F^0_{c+N,d+N}(q_c-q_d)\Big)+
 }
 \\ \ \\
  \displaystyle{
 +\frac{1}{2}\sum\limits_{c \neq d}
 \Big(F^0_{c,d+N}(q_c+q_d)+F^0_{c+N,d}(q_c+q_d)) + \frac{\mu}{\nu}\sum\limits_{c}(F^0_{c,c+N}(2q_c)+F^0_{c+N,c}(2q_c)\Big)
 +}
 \\ \ \\
 \displaystyle{
 +\sum\limits_{c} F^0_{c,2N+1}(q_c)+\sum\limits_{c}
 F^0_{c+N,2N+1}(q_c) \,.
 }
 \end{array}
 \eq
After rearranging the summands we obtain:
  \beq\label{a386}
  \begin{array}{c}
   \displaystyle{
\nu g \Big[R^{z}_{a,2N+1}(q_a), \F^0 - \sum_{b: \, b \neq
a}(F^{0}_{ab}(q_a-q_b) + F^{0}_{a,b+N}(q_a+q_b)) -
 }
 \\ \ \\
  \displaystyle{
 -
\frac{\mu}{\nu}F^0_{a,a+N}(2q_a) -
\sum\limits_{c}F^0_{c,2N+1}(q_c)\Big] +
 }
 \\ \ \\
    \displaystyle{
 +\Big(\frac{g^3}{\nu} \, F^0_{a,2N+1}(q_a)  - \nu g \,
F^0_{a,2N+1}(q_a) +\mu g\, F^0_{a+N,2N+1}(q_a) - \nu g \,
F^0_{a+N,2N+1}(q_a)\Big)\times
 }
 \\ \ \\
    \displaystyle{
\times R^{z}_{a,2N+1}(q_a)\,.
    }
 \end{array}
 \eq
The commutator vanishes because all terms, which act non-trivially
in $a$ and $2N+1$ quantum spaces are subtracted from $\F^0$, and we
are left with the following expression:
  \beq\label{a387}
  \begin{array}{c}
   \displaystyle{
\Big(\frac{g^3}{\nu}-\nu g\Big) F^0_{a,2N+1}(q_a) + (\mu g-\nu g)
F^0_{a+N,2N+1}(q_a) }
 \end{array}
 \eq
(multiplied by $R^z_{a,2N+1}(q_a)$), which must be equal to zero for
validity of the Lax equations. In the scalar case ($\tilde{N}=1$)
$F^0_{b,2N+1}$ is proportional to $F^0_{b+N,2N+1}$,
   and this requirement leads only to the condition (\ref{a102}).
 But in the case  $\tilde{N} \neq 1$ it totally fixates the constants (up to a sign).
 Calculations for blocks \{23\}, \{31\} and \{32\} are absolutely
 analogous. And all other blocks in $[\mL,\mM]$ depend only on square of $g$, consequently are not sensitive to its sign.
 The natural way to solve such kind of problem is to reduce the number of
 quantum spaces. This is what we do in the next paragraph.

\subsection{${\rm D}_{N}$ case with $N$ quantum spaces}

$R$-matrix-valued ${\rm SO}_{2N}$ Lax pair with $N$ quantum spaces
has the same block structure as the one for ${\rm Sp}(2N)$ with $2N$
quantum spaces (\ref{SO(2N)}):
 \begin{equation}\label{a47}
\mathcal{L} =
     \left(\begin{array}{cc}
    P+A_1 & B_1           \\
    B_2 & -P+A_2
         \end{array}
         \right)
\quad\quad
 \mathcal{M} =
      \left(\begin{array}{cc}
    A_1'+D_1+\mathcal{F} & B_1'          \\
    B_2' & A_2'+D_2+\mathcal{F}
         \end{array}\right)
 \end{equation}
 But
now the corresponding blocks have the form:
 \beq\label{a471}\begin{array}{c}
(A_1)_{ab} = \nu (1-\delta_{ab})\,
R^{\,z}_{ab}(q_a-q_b)\,,\quad\quad
(A_2)_{ab} = \nu (1-\delta_{ab}) \, R^{\,z}_{ab}(-q_a+q_b)\,,\\ \ \\
(B_1)_{ab} = \nu (1-\delta_{ab}) \,
R^{\,z}_{ab}(q_a+q_b)\,,\quad\quad
(B_2)_{ab} = \nu (1-\delta_{ab}) \, R^{\,z}_{ab}(-q_a-q_b)\,,\\ \ \\
(A_1')_{ab} = \nu (1-\delta_{ab}) \,
F^{\,z}_{ab}(q_a-q_b)\,,\quad\quad
(A_2')_{ab} = \nu (1-\delta_{ab}) \, F^{\,z}_{ab}(-q_a+q_b)\,,\\ \ \\
(B_1')_{ab} = \nu (1-\delta_{ab}) \,
F^{\,z}_{ab}(q_a+q_b)\,,\quad\quad
(B_2')_{ab} = \nu (1-\delta_{ab}) \, F^{\,z}_{ab}(-q_a-q_b)\,,\\ \ \\
(D_1)_a = (D_2)_a =  D_a = -\nu \sum\limits_{{c:\, c \ne a}}
(F^0_{ac}(q_a-q_c) + F^0_{ac}(q_a+q_c))\,,
 \end{array}\eq
 \beq\label{a40}\begin{array}{c}
\mathcal{F}_{ab}= \nu  \delta_{ab} \F^0\,,\\ \ \\
\displaystyle{ \F^0 = \frac{1}{2}\sum\limits_{c \neq d}
(F^0_{cd}(q_c-q_d)+F^0_{cd}(q_c+q_d))\,.}
 \end{array}\eq
Notice that blocks $B_1$ and $B_2$ are off-diagonal here.
Representing the Lax pair schematically as $\mL=\mathcal{P}+R$,
$\mM=D+\F+F$, where $\mathcal{P}$ consists of momenta part, $R$
includes $A_{1},A_2,B_{1},B_2$ and similarly for $\mM$, we can rewrite the r.h.s.
of the Lax equations in the form:
 \beq\label{a41}\begin{array}{c}
[\mL,\mM] = [\mathcal{P}+R, D+\F+F] = [\mathcal{P},F] + [R,D] + [R,
\F] + [R,F]\,.
 \end{array}\eq
The calculations are performed for $N\times N$ ($\otimes {\rm
Mat}_{\ti N}^{\otimes N}$) blocks \{11\}, \{12\}, \{21\}, \{22\}
separately.

\subsubsection*{Proof for Block \{11\}}
The first summand in (\ref{a41}):
 \beq\begin{array}{c}
[P,F] = \sum\limits_{c, a \neq b} \nu p_c [E_{cc} \ox 1, E_{ab} \ox
F^{\,z}_{ab}(q_a-q_b)] = \sum\limits_{a \neq b} \nu
(\dot{q}_a-\dot{q}_b) E_{ab} \ox F^{\,z}_{ab}(q_a-q_b)\,.
 \end{array}\eq
The second summand in (\ref{a41}):
 \beq\begin{array}{c}
[R,D] = \sum\limits_{a \neq b} \nu^2 E_{ab} \ox (R^{\,z}_{ab}(q_a-q_b) D_b - D_a R^{\,z}_{ab}(q_a-q_b)) = \\
 \sum\limits_{a \ne b} \nu^2
  E_{ab}\otimes\Big(\sum\limits_{{c:\, c \ne a}} F^0_{ac}(q_a-q_c)R^{\,z}_{ab}(q_a-q_b) + F^0_{ac}(q_a+q_c)R^{\,z}_{ab}(q_a-q_b) - \\
 - \sum\limits_{{c:\, c \ne b}} R^{\,z}_{ab}(q_a-q_b)F^0_{bc}(q_b-q_c) +
 R^{\,z}_{ab}(q_a-q_b)F^0_{bc}(q_b+q_c)\Big)=\\
 =
 \sum\limits_{a \neq b} \nu^2 E_{ab} \ox \sum\limits_{{c:\, c \ne
a,b}}\Big(F^0_{ac}(q_a-q_c)R^{\,z}_{ab}(q_a-q_b) +
F^0_{ac}(q_a+q_c)R^{\,z}_{ab}(q_a-q_b) - \\ -
R^{\,z}_{ab}(q_a-q_b)F^0_{bc}(q_b-q_c) -
R^{\,z}_{ab}(q_a-q_b)F^0_{bc}(q_b+q_c))
+F^0_{ab}(q_a-q_b)R^{\,z}_{ab}(q_a-q_b) +\\+
F^0_{ab}(q_a+q_b)R^{\,z}_{ab}(q_a-q_b) -
R^{\,z}_{ab}(q_a-q_b)F^0_{ba}(q_b-q_a) -
R^{\,z}_{ab}(q_a-q_b)F^0_{ba}(q_b+q_a)\Big)\,.
 \end{array}\eq
The last summand in (\ref{a41}):
 \beq\begin{array}{c}
[R,F]\hbox{(off-diagonal)}=\\ \ \\ =  \sum\limits_{a \neq b} \nu^2
E_{ab} \ox \sum\limits_{{c: \, c \ne a,b}} \Big(
R^{\,z}_{ac}(q_a-q_c)F^{\,z}_{cb}(q_c-q_b) -
F^{\,z}_{ac}(q_a-q_c)R^{\,z}_{cb}(q_c-q_b) +\\+
R^{\,z}_{ac}(q_a+q_c)F^{\,z}_{cb}(-q_c-q_b)-
F^{\,z}_{ac}(q_a+q_c)R^{\,z}_{cb}(-q_c-q_b) \Big)
\stackrel{(\ref{a17})}{=}\\ \ \\=\sum\limits_{a\ne b} \nu^2 E_{ab}
\ox \sum\limits_{{c : \,  c \ne a,b}}\Big(
F^0_{cb}(q_c-q_b)R^{\,z}_{ab}(q_a-q_b)-
R^{\,z}_{ab}(q_a-q_b)F^0_{ac}(q_a-q_c)+\\+
F^0_{cb}(-q_c-q_b)R^{\,z}_{ab}(q_a-q_b) -
R^{\,z}_{ab}(q_a-q_b)F^0_{ac}(q_a+q_c)\Big)\,.
 \end{array}\eq
 \beq\begin{array}{c}
[R,F]\hbox{(diagonal)} =\\ \ \\ =\sum\limits_a \nu^2 E_{aa} \ox
\sum\limits_{{c: \, c \ne a}}\Big(
R^{\,z}_{ac}(q_a-q_c)F^{\,z}_{ca}(q_c-q_a) -
F^{\,z}_{ac}(q_a-q_c)R^{\,z}_{ca}(q_c-q_a) +\\+
R^{\,z}_{ac}(q_a+q_c)F^{\,z}_{ca}(-q_c-q_a)-
F^{\,z}_{ac}(q_a+q_c)R^{\,z}_{ca}(-q_c-q_a)\Big)\stackrel{(\ref{a192})}{=}\\
\ \\
 =
 \sum\limits_a \nu^2 \tilde{N}^2 E_{aa} \ox
\sum\limits_{{c: \, c \ne a}} (\wp'(q_a-q_b) + \wp'(q_a+q_b))\,.
 \end{array}\eq
 Hence
 \beq\begin{array}{c}
\dot{\mL}= [P,F] +[R,F]\hbox{(diagonal)}
  \end{array}\eq
 provides equations of motion (with $\nu$ replaced by $\tilde{N} \nu$).
Let us verify that the sum of the remaining terms in the r.h.s. of the
Lax equations is equal to zero. Indeed,
 \beq\label{a141}\begin{array}{c}
[R,F]\hbox{(off-diagonal)}+[R,\F]+[R,D] =\\ \ \\=  \sum\limits_{a
\neq b} \nu^2 E_{ab} \ox \Big[R^{\,z}_{ab}(q_a-q_b),\F^0 -
F^0_{ab}(q_a-q_b) - F^0_{ab}(q_a+q_b) -\qquad\qquad\\
\qquad\qquad-\sum\limits_{{c: \, c \ne a,b}} (F^0_{ac}(q_a-q_c) +
F^0_{ac}(q_a+q_c) + F^0_{bc}(q_b-q_c) + F^0_{bc}(q_b+q_c))\Big]\,.
 \end{array}\eq
This equals zero, since there are no terms in the right part of the
commutator acting non-trivially in $a$-th and $b$-th quantum spaces.
All such terms are cancelled by $\F^0$.

\subsubsection*{Proof for block \{12\}}
The r.h.s. of the Lax equations:
 \beq\label{a42}\begin{array}{c}
[L,M]_{a,b+N} = [\mathcal{P}+R, D+\F+F]_{a,b+N} =\\ \ \\=
[\mathcal{P},F]_{a,b+N} + [R,D]_{a,b+N} + [R, \F]_{a,b+N} +
[R,F]_{a,b+N},
 \end{array}\eq
As in the case of block \{12\}  the term $[\mathcal{P},F]_{a,b+N}$
is cancelled by the off-diagonal part in the block  \{12\} of the
l.h.s. of the Lax equation. But \{12\} block of the Lax operator has
no diagonal part in this new case. So the sum of the remaining terms
in (\ref{a42}) should vanish:
 \beq\label{a43}\begin{array}{c}
[R,D]_{a,b+N} + [R, \F]_{a,b+N} + [R,F]_{a,b+N}=0
 \end{array}\eq
Let us verify it. First, consider the last term in (\ref{a43}):
 \beq\begin{array}{c}
[R,F]_{a,b+N}\hbox{(off-diagonal)} =\\ \ \\= \nu^2 \sum\limits_{{c:
\, c \ne a,b}} \Big( R^{\,z}_{ac}(q_a-q_c)F^{\,z}_{cb}(q_c+q_b) -
F^{\,z}_{ac}(q_a-q_c)R^{\,z}_{cb}(q_c+q_b) +\\+
R^{\,z}_{ac}(q_a+q_c)F^{\,z}_{cb}(-q_c+q_b)-
F^{\,z}_{ac}(q_a+q_c)R^{\,z}_{cb}(-q_c+q_b)\Big)\stackrel{(\ref{a17})}{=}
 \\ \ \\
  = \nu^2 \sum\limits_{{c : \, c
\ne a,b}} \Big( F^0_{cb}(q_c+q_b)R^{\,z}_{ab}(q_a+q_b)-
R^{\,z}_{ab}(q_a+q_b)F^0_{ac}(q_a-q_c)+\\+
F^0_{cb}(-q_c+q_b)R^{\,z}_{ab}(q_a+q_b) -
R^{\,z}_{ab}(q_a+q_b)F^0_{ac}(q_a+q_c)\Big)\,.
 \end{array}\eq
 For the diagonal part there is no appropriate identity, and we leave it as it is:
 \beq\begin{array}{c}
[R,F]_{a,a+N}\hbox{(diagonal)} =\\ \ \\ =\nu^2 \sum\limits_{{c : \,
c \ne a}}\Big( R^{\,z}_{ac}(q_a-q_c)F^{\,z}_{ca}(q_c+q_a) -
F^{\,z}_{ac}(q_a-q_c)R^{\,z}_{ca}(q_c+q_a) +\\+
R^{\,z}_{ac}(q_a+q_c)F^{\,z}_{ca}(-q_c+q_a)-
F^{\,z}_{ac}(q_a+q_c)R^{\,z}_{ca}(-q_c+q_a)\Big)\,.
 \end{array}\eq
The first term in (\ref{a43}) is again simplified through
(\ref{a17}):
 \beq\begin{array}{c}
 [R,D]_{a,b+N} = (R^{\,z}_{ab}(q_a-q_b) D_b - D_a
R^{\,z}_{ab}(q_a-q_b)) = \\ \ \\
 = \nu^2 \sum\limits_{{c : \, c \ne
a,b}}\Big(F^0_{ac}(q_a-q_c)R^{\,z}_{ab}(q_a+q_b) +
F^0_{ac}(q_a+q_c)R^{\,z}_{ab}(q_a+q_b) - \\ \ \\ -
R^{\,z}_{ab}(q_a+q_b)F^0_{bc}(q_b-q_c) -
R^{\,z}_{ab}(q_a+q_b)F^0_{bc}(q_b+q_c) + \\ \ \\
+F^0_{ab}(q_a-q_b)R^{\,z}_{ab}(q_a+q_b) +
F^0_{ab}(q_a+q_b)R^{\,z}_{ab}(q_a+q_b) - \\ \ \\ -
R^{\,z}_{ab}(q_a+q_b)F^0_{ba}(q_b-q_a) -
R^{\,z}_{ab}(q_a+q_b)F^0_{ba}(q_b+q_a)\Big)
 \end{array}\eq
Gathering all the off-diagonal terms in (\ref{a43}) we get the
commutator
 \beq\begin{array}{c}
[R,F]_{a,b+N}\hbox{(off-diagonal)} + [R,D]_{a,b+N} + [R, \F]_{a,b+N}
= \\ \ \\ = \nu^2 \Big[R^{\,z}_{ab}(q_a+q_b),\F^0 -
F^0_{ab}(q_a-q_b) - F^0_{ab}(q_a+q_b) -\\ \ \\ -\sum\limits_{{c : \,
c \ne a,b}} \Big(F^0_{ac}(q_a-q_c) + F^0_{ac}(q_a+q_c) +
F^0_{bc}(q_b-q_c) + F^0_{bc}(q_b+q_c)\Big)\Big]\,,
 \end{array}\eq
which equals zero by the same reason as for block \{11\}
(\ref{a141}). Therefore, we are left with the diagonal term
 \beq\begin{array}{c}
[\mL,\mM]_{a,a+N} = [R,F]_{a,a+N}\,,
 \end{array}\eq
 and it vanishes if
 \begin{equation}
\label{RR} R^{\,z}_{ab}(u)F^{\,z}_{ba}(v) -
F^{\,z}_{ab}(v)R^{\,z}_{ba}(u) = 0\,.
 \end{equation}
 The latter is of course not true in general case, but is true in some particular cases. For example, it is obviously holds true
 for the Yang's case (\ref{a13}). What is more important for us is that (\ref{RR}) holds true in $\ti N=2$ case
  with the Baxter's $R$-matrix since it is of the form $R^{\,z}_{12}(u) = \sum\limits_{\alpha = 0}^{3} \vf^z_\alpha(u) \;
\sigma_\alpha \ox \sigma_\alpha$. It is easy to check that for such
$R$-matrices
 \beq\label{a45}\begin{array}{c}
[R^{\,z}_{ab}(u), R^{\,z}_{ab}(v)] = 0
 \end{array}\eq
due to the properties of the Pauli matrices. By differentiating
(\ref{a45}) with respect to $v$ (and using the additional symmetry
$R^{\,z}_{ab}(u)=R^{\,z}_{ba}(u)$) one finds (\ref{RR}). With this
property we have
 \beq\begin{array}{c}
R^{\,z}_{ac}(q_a-q_c)F^{\,z}_{ca}(q_c+q_a) -
F^{\,z}_{ac}(q_a-q_c)R^{\,z}_{ca}(q_c+q_a) +\\ \ \\+
R^{\,z}_{ac}(q_a+q_c)F^{\,z}_{ca}(-q_c+q_a) -
F^{\,z}_{ac}(q_a+q_c)R^{\,z}_{ca}(-q_c+q_a) = 0\,,
 \end{array}\eq
and this is the expression, standing under the sum in
$[R,F]_{a,a+N}$.

The proofs for blocks \{21\} and \{22\} are performed in a similar
way.

\subsection{${\rm B}_{N}$ case with $N+1$ quantum spaces}
Here $R$-matrix-valued Lax pair with $N+1$ quantum spaces has the
same block structure as the one for ${\rm BC}(N)$ with $2N+1$
quantum spaces (\ref{BC(N)})
 \begin{equation}
\label{a48}
 \mathcal{L} =
     \left(\begin{array}{ccc}
    P+A_1 & B_1 & C_1     \\
    B_2 & -P+A_2 & C_2    \\
    C_2^T & C_1^T & 0
         \end{array}\right)
 \quad
\mathcal{M} =
     \left(\begin{array}{ccc}
    A_1'+D_1+\mathcal{F} & B_1' & C_1'    \\
    B_2' & A_2'+D_2+\mathcal{F} & C_2' \\
    C_2'^T & C_1'^T & D_3+\mathcal{F}
         \end{array}\right)
 \end{equation}
but now the corresponding blocks have the following form.
The blocks $A$, $B$ and $P$ are the same as in the previous case
(\ref{a471}), and the rest
are:
 \beq\begin{array}{c}
(C_1)_a = \pm\sqrt{2}\nu \,  R^{\,z}_{a,N+1}(q_a)\,,\quad (C_2)_a =
\pm\sqrt{2}\nu  \,R^{\,z}_{a,N+1}(-q_a)\,,\\ \ \\
(C_1^T)_a =
\pm\sqrt{2}\nu \, R^{\,z}_{N+1,a}(q_a)\,,\quad (C_2^T)_a =
\pm\sqrt{2}\nu \, R^{\,z}_{N+1,a}(-q_a)\,,\\ \ \\
(C_1')_a = \pm\sqrt{2}\nu \, F^{\,z}_{a,N+1}(q_a)\,,\quad (C_2')_a =
\pm\sqrt{2}\nu \, F^{\,z}_{a,N+1}(-q_a)\,,\\ \ \\
(C_1'^T)_a = \pm\sqrt{2}\nu \, F^{\,z}_{N+1,a}(q_a)\,,\quad
(C_2'^T)_a =
\pm\sqrt{2}\nu \, F^{\,z}_{N+1,a}(-q_a)\,,\\ \ \\
 D_3 =
-2\nu\; \sum\limits_{c} F^0_{c,N+1}(q_c)\,.
 \end{array}\eq
The $D$ and $\F$ terms are given by:
 \beq\label{a49}\begin{array}{c}
(D_1)_a = (D_2)_a = -\nu \sum\limits_{{c:\, c \ne a}}
 \Big(F^0_{ac}(q_a-q_c) + F^0_{ac}(q_a+q_c)\Big)- 2 \nu \;F^0_{a,N+1}(q_a)\,,
 \end{array}\eq
and
 \beq\label{a50}\begin{array}{c}
\mathcal{F}_{ab}= \nu  \delta_{ab} \F^0\,,\quad\quad \displaystyle{
\F^0 = \frac{1}{2}\sum\limits_{c \neq d}
 \Big(F^0_{cd}(q_c-q_d)+F^0_{cd}(q_c+q_d)\Big)+2 \sum\limits_{c}
F^0_{c,N+1}(q_c)\,.}
 \end{array}\eq
Or, equivalently
 \beq\begin{array}{c}
(D_1)_a^{BC(N)} = (D_2)_a^{BC(N)} = D_a^{SO(2N)} - 2 \nu
\;F^0_{a,N+1}(q_a)\,,\\ \ \\
 \nu \, \F^0_{BC(N)} = \nu \,
\F^0_{SO(2N)} - D_3\,.
 \end{array}\eq
 As in the previous case the calculations are performed separately
 for $N\times N$ ($\otimes {\rm
Mat}_{\ti N}^{\otimes N}$) blocks \{11\}, \{12\}, \{21\}, \{22\};
for $N\times 1$ ($\otimes {\rm Mat}_{\ti N}^{\otimes N}$) blocks
\{13\}, \{23\}; for $1\times N$ ($\otimes {\rm Mat}_{\ti N}^{\otimes
N}$) blocks \{31\}, \{32\} and for $1\times 1$ ($\otimes {\rm
Mat}_{\ti N}^{\otimes N}$) block \{33\}. Evaluations for the blocks
\{11\}, \{12\}, \{21\}, \{22\} are similar. In particular, we again
come to condition (\ref{RR}). The calculation for the block \{33\}
is as follows:
 \beq\begin{array}{c}
[L,M]_{2N+1,2N+1} = (C_2^TC_1'- C_2'^TC_1 + C_1^TC_2'- C_1'^TC_2)=
 \\ \ \\
 = 2\nu^2 \sum\limits_a\Big(
R^{\,z}_{N+1,a}(-q_a)F^{\,z}_{a,N+1}(q_a)-F^{\,z}_{N+1,a}(-q_a)R^{\,z}_{a,N+1}(q_a)
+\\+ R^{\,z}_{N+1,a}(q_a)F^{\,z}_{a,N+1}(-q_a) -
F^{\,z}_{N+1,a}(q_a)R^{\,z}_{a,N+1}(-q_a)\Big)\stackrel{(\ref{a192})}{=}
 \\ \ \\
 =2\nu^2 \sum\limits_a(\tilde{N}^2 \wp'(-q_a) +
\tilde{N}^2 \wp'(q_a)) = 0\,.
 \end{array}\eq

In the end of the Section let us remark that the Hamiltonian
(\ref{a10}) is a particular case of the most general model
\cite{Inoz89}
 \begin{equation}
 \label{a101}
 H = \frac{1}{2} \sum_{a=1}^{N} p_a^2 - \nu^2
\sum_{a<b}^{N} (\wp(q_a-q_b) + \wp(q_a+q_b)) -
\sum\limits_{\ga=0}^3\sum_{a=1}^{N} \nu_\ga^2\wp(q_a+\om_\ga)\,,
 \end{equation}
which has $3N\times 3N$ Lax representation (and $2N\times 2N$ Lax
representation \cite{Feher2}). In the last term $\om_\ga$ are the
half-periods of the elliptic curve with moduli $\tau$:
$\{0,1/2,\tau/2,1/2+\tau/2\}$, and all five constants in
(\ref{a101}) are arbitrary.
Unfortunately, we do not know how to extend $3N\times 3N$ (or
$2N\times 2N$) Lax pair for (\ref{a101}) to the $R$-matrix-valued
case. At the same time the ${BC}_1$ case (one degree of freedom and
four arbitrary constants) with $N=2$ was suggested in \cite{LOZ15}.



\section{Quantum Lax pairs and spin models}\label{quant}
\setcounter{equation}{0}

\paragraph{Spinless case.} The classical Lax pairs for Calogero-Moser models
can be quantized through the quantum Lax equation \cite{Ujino,BMS}
  \beq\label{a60}
  \begin{array}{c}
  \displaystyle{
[\hat{H},\hat{L}(z)]=\hbar\,[\hat{L}(z),M(z)]\,.
 }
 \end{array}
 \eq
 Before
discussing the $R$-matrix-valued Lax pairs consider the ordinary
case $\ti N=1$. As mentioned before, in this case the Lax pair
(\ref{a07})-(\ref{a08}) differs from (\ref{a04})-(\ref{a05}) by the
scalar (in the auxiliary space) part $\mF^0$ (\ref{a09}) of
$M$-matrix. For $\ti N=1$
  \beq\label{a61}
  \begin{array}{c}
    \displaystyle{
\mF^0|_{\ti
N=1}=-\nu\sum\limits_{i>j}\wp(q_{ij})+\hbox{const}\,,\quad\quad\hbox{const}=\frac{N^2-N}{3}\frac{\vth'''(0)}{\vth'(0)}\,.
 }
 \end{array}
 \eq
 This term is cancelled out from the classical Lax equations. But it
 becomes important in quantum case because it is treated as a part
 of the quantum Hamiltonian in (\ref{a60}). With this term the coupling
 constant in
  \beq\label{a62}
  \begin{array}{c}
    \displaystyle{
 \hat{H}=\frac{1}{2}\sum\limits_{i=1}^N{\hat
 p}_i^2-\nu(\nu+\hbar\,)\sum\limits_{i>j}^N\wp(q_i-q_j)\,,\quad\quad
 {\hat p}_i=\hbar\,\p_{q_i}
 }
 \end{array}
 \eq
 acquires the quantum correction. Another reason for treating
 (\ref{a61}) as a part of the Hamiltonian is as follows. After
 subtracting $\mF^0$ from $M$ we are left (in $\ti N=1$ case) with the
 original
 $M$-matrix (\ref{a05}). In rational and trigonometric cases
 (\ref{a908}) this $M$-matrix satisfies the so-called sum up to zero
 condition:
  \beq\label{a63}
  \begin{array}{c}
    \displaystyle{
 \sum\limits_{j\,:j\neq i} M_{ij}=\sum\limits_{j\,:j\neq i}
 M_{ji}=0\quad\hbox{for}\ \hbox{all}\ i\,.
 }
 \end{array}
 \eq
It can be proved \cite{Ujino} that with this condition the total
sums
  \beq\label{a64}
  \begin{array}{c}
    \displaystyle{
 {\hat I}_k={\rm ts}({\hat L}^k)=\sum\limits_{i,j} ({\hat L}^k)_{ij}\,,\quad\quad
 k\in\mZ_+
 }
 \end{array}
 \eq
are the quantum integrals of motion, i.e. $[\hat{H},\hat{I}_k]=0$
(while in the classical case the integrals of motion can be defined
as $\tr(L^k)$). The second Hamiltonian ${\hat I}_2$ yields $\hat{H}$
(\ref{a62}).

 In the elliptic case the sum up to zero
 condition (\ref{a63}) is not fulfilled. At the same time the quantum Lax
 equation holds true.

\paragraph{Spin case.} Similar construction works for the spin
Calogero-Moser model \cite{GH}. The operator valued Lax pair
\cite{Inoz0,HW,Bernard,Inoz1} is of the form
  \beq\label{a65}
  \begin{array}{c}
  \displaystyle{
{\hat L}^{\hbox{\tiny{spin}}}(z)=\sum\limits_{i,j=1}^N
E_{ij}\otimes{\hat L}^{\hbox{\tiny{spin}}}_{ij}(z)\,,\quad {\hat
L}^{\hbox{\tiny{spin}}}_{ij}(z)=\delta_{ij}p_i+\nu(1-\delta_{ij})\phi(z,q_{ij})P_{ij}\,,
 }
 \end{array}
 \eq
  \beq\label{a66}
  \begin{array}{c}
  \displaystyle{
{\hat M}^{\hbox{\tiny{spin}}}_{ij}(z)=\nu d_i\delta_{ij}
+\nu(1-\delta_{ij})f(z,q_{ij})P_{ij}\,,\quad d_i=\sum\limits_{k:
k\neq i}^N E_2(q_{ik})P_{ik}=-\sum\limits_{k: k\neq i}^N
f(0,q_{ik})P_{ik}\,,
 }
 \end{array}
 \eq
where $P_{ij}\in{\rm Mat}_{\ti N}^{\otimes N}$ is the permutation
operator (of $i$-th and $j$-th tensor components)
 in quantum space  ${\mathcal
H}=(\mC^{\ti N})^{\otimes N}$, or the spin exchange operator. The
tensor structure
 of (\ref{a65})-(\ref{a66}) is the same as for the $R$-matrix-valued
 Lax pair (\ref{a07})-(\ref{a08}). The term $\mF^0$ (\ref{a08}) in this case
  \beq\label{a67}
  \begin{array}{c}
    \displaystyle{
 \mF^0=-\nu\sum\limits_{i>j}E_2(q_i-q_j)P_{ij}
 }
 \end{array}
 \eq
 is
 treated as a part of the quantum Hamiltonian
  \beq\label{a68}
  \begin{array}{c}
    \displaystyle{
 \hat{H}^{\hbox{\tiny{spin}}}=\frac{1}{2}\sum\limits_{i=1}^N{\hat
 p}_i^2-\sum\limits_{i>j}^N \nu(\nu+\hbar\, P_{ij})E_2(q_i-q_j)
 }
 \end{array}
 \eq
likewise it was performed in the spinless case (\ref{a62}). It
describes the spin exchange interaction. Again, in the rational and
trigonometric cases the sum up to zero condition is fulfilled, and
${\rm ts}({\hat L}^k)$ provides the higher Hamiltonians including
the one (\ref{a68}) for $k=2$. Another recipe for constructing the
higher Hamiltonians comes from the underlying Yangian structure
\cite{Bernard}.

From the point of view of $R$-matrix-valued case
(\ref{a07})-(\ref{a08}) the Lax pair ${\hat
L}^{\hbox{\tiny{spin}}}(z)$ and ${\hat
M}^{\hbox{\tiny{spin}}}(z)+1_N\otimes\mF^0$ corresponds to the
special case when $R_{ij}^z(q_{ij})$ is replaced by
$\phi(z,q_{ij})P_{ij}$. The associative Yang-Baxter equation
(\ref{a16}) holds true in this case since the matrix-valued
expressions $P_{ab}P_{bc}=P_{ac}P_{ab}=P_{bc}P_{ac}$ are cancelled
out, and we are left with the scalar identity (\ref{a14}).

\paragraph{Anisotropic spin case.} In \cite{HW2} the anisotropic
extension of the (trigonometric) spin Calogero-Moser-Sutherland
model was suggested. It is based on the ${\rm su}(2)$ XXZ
(trigonometric) classical $r$-matrix
  \beq\label{a69}
  \begin{array}{c}
    \displaystyle{
 r_{12}^{\hbox{\tiny{XXZ}}}(q)=
 (1\otimes1+\sigma_3\otimes\sigma_3)\cot(q)+(\sigma_1\otimes\sigma_1+\sigma_2\otimes\sigma_2)\frac{1}{\sin(q)}\,,
 }
 \end{array}
 \eq
 which is used in the Lax pair
in the same way as $R_{ij}^z(q)$ in (\ref{a07}):
  \beq\label{a70}
  \begin{array}{c}
  \displaystyle{
{\hat L}^{\hbox{\tiny{XXZ}}}(z)=\sum\limits_{i,j=1}^N
E_{ij}\otimes{\hat L}^{\hbox{\tiny{XXZ}}}_{ij}(z)\,,\quad {\hat
L}^{\hbox{\tiny{XXZ}}}_{ij}(z)=\delta_{ij}p_i+\nu(1-\delta_{ij})r^{\hbox{\tiny{XXZ}}}_{ij}(q_{ij})\,.
 }
 \end{array}
 \eq
 In this respect the
Lax matrix (\ref{a07}) with the elliptic Baxter-Belavin $R$-matrix
generalizes the case (\ref{a69}) in the same way as the elliptic Lax
matrix (\ref{a04}) generalizes the one without spectral parameter
for the classical (trigonometric) Sutherland model. Let us slightly
modify (\ref{a70}) to have exactly the form (\ref{a07}). Consider
(\ref{a07})-(\ref{a08}) defined through the XXZ quantum $R$-matrix
  \beq\label{a702}
  \begin{array}{c}
    \displaystyle{
 R^{\,z}_{12}(q) = \sum_{\alpha = 0 }^3 \vf^z_\alpha(q)\; \sigma_\alpha \ox
 \sigma_\alpha\,,\quad\quad \vf^z_1(q) = \vf^z_2(q) = \frac{1}{\sin{q}}\,,
 }
 \\
     \displaystyle{
 \vf^z_0(q) =
 (\cot{z}+\cot{q}+\frac{1}{\sin{z}})\,,\quad\quad\ \vf^z_3(q) =
 (\cot{z}+\cot{q}-\frac{1}{\sin{z}})\,.
 }
 \end{array}
 \eq
This results in the quantum Hamiltonian with the $\mF^0$ term
  \beq\label{a703}
  \begin{array}{c}
  \displaystyle{
\mF^0=\sum_{a < b}
\Big(\frac{1}{\sin^2{q_{ab}}}(\stackrel{a}{\sigma}_0 \ox
\stackrel{b}{\sigma}_0 + \stackrel{a}{\sigma}_3 \ox
\stackrel{b}{\sigma}_3) +
\frac{\cos{q_{ab}}}{\sin^2{q_{ab}}}(\stackrel{a}{\sigma}_1 \ox
\stackrel{b}{\sigma}_1 + \stackrel{a}{\sigma}_2 \ox
\stackrel{b}{\sigma}_2)\Big)\,,
 }
 \end{array}
 \eq
which was obtained in \cite{HW2} from (\ref{a70}) via (\ref{a64}).
It happened because the terms depending on $z$ in (\ref{a702}), do
not depend on $q$. Hence, $F^z_{12}(q)=F^0_{12}(q)$ and the
condition (\ref{a63}) can be fulfilled.

For the elliptic $R$-matrix-valued Lax pairs with half amount of
quantum spaces, found in the previous Section, the quantum Lax
equations hold true in a similar way:
\begin{predl}
For the Lax pairs (\ref{a07})-(\ref{a08}) for ${\rm A}_{N-1}$,
(\ref{a47}) for ${\rm D}_{N}$ and (\ref{a48}) for ${\rm B}_{N}$ root
systems one can define the quantum Lax pair by simple replacing
momenta $p_i$ with ${\hat p}_i=\hbar\,\p_{q_i}$. The quantum Lax
equations hold true with the quantum Hamiltonian ${\hat H}$ obtained
from the classical $H(p,q)$ as $H({\hat p},q)$.
\end{predl}
The proof of this statement is direct. It is similar to the
classical case and (\ref{a65}). The main idea, needed for the proof
is that in the above mentioned cases
  \beq\label{a700}
  \begin{array}{c}
  \displaystyle{
[P,D+\F^0] = 0\,,
 }
 \end{array}
 \eq
where $D$ is a  diagonal part of the $M$-operator. This happens
because the only terms in $D$ depending on some $q_a$ are those,
which simultaneously act non-trivially on the tensor component
number $a$ in quantum space. We already know from the classical case
that $D_a$ subtracts exactly these terms from $\F^0$.
 For ${\rm C}_{N}$ and ${\rm BC}_{N}$ cases this statement is not true.
 For example, $D_{1a} - \F^0$ contains the terms $F^0_{a+N,c+N}(-q_a+q_c)$, which do not commute with
 $\hat{p}_a$. $\blacksquare$

From what has been said it seems also natural to remove the $\mF^0$
terms from the corresponding $M$-matrices and treat them as a part
of the quantum Hamiltonians, which then take the form
  \beq\label{a71}
  \begin{array}{c}
  \displaystyle{
 {\hat H}=H(\hat{p},q)+\hbar\,\nu\mF^0\,,
 }
 \end{array}
 \eq
where the terms $\mF^0$ are given by (\ref{a09}) for ${\rm
A}_{N-1}$, (\ref{a47}) for ${\rm D}_{N}$ and (\ref{a50}) for ${\rm
B}_{N}$ root systems.

Let us emphasize that $\mF^0$ term in the elliptic quantum
Hamiltonian (\ref{a71}) does not follow from the Lax matrix. The
construction (\ref{a64}) used in the rational and trigonometric
cases does not work here since the sum up to zero condition
(\ref{a63}) is not valid. Moreover, in the classical case the
corresponding Lax matrices provide the Hamiltonians without $\mF^0$.
However, we are going to argue that (\ref{a71}) is meaningful even
in the classical case (see the next Section).

The quantum Lax matrices (\ref{a65}),  are closely related to the
Knizhnik-Zamolodchikov equations \cite{KZ}. As was mentioned in
\cite{HW,HW2} the KZ connections (\ref{a70}) are obtained as
  \beq\label{a72}
  \begin{array}{c}
  \displaystyle{
\nabla_i=\sum\limits_{j=1}^N {\hat
L}^{\hbox{\tiny{XXZ}}}_{ij}(z)=\hbar\,\p_{q_i}+\nu\sum\limits_{j:\,j\neq
i}^N r_{ij}(q_{ij})\,,
 }
 \end{array}
 \eq
%
and the ground state for this model solves the KZ equations
  \beq\label{a73}
  \begin{array}{c}
  \displaystyle{
\nabla_i\psi=0\,,\quad i=1...N\,.
 }
 \end{array}
 \eq
The interrelation between the quantum Calogero models and the KZ
equations was observed by Matsuo and Cherednik \cite{Matsuo,Chered0}
(see also\cite{FV}). Their construction, in fact, underlies the
construction of the operator valued Lax pairs.\footnote{ It should
be also mentioned that different types of spinless and spin
Calogero-Moser models can be described in the framework of the
Cherednik-Dunkl and/or exchange operator formalism
\cite{CheredDunkl,Polych0}. See e.g. \cite{Bernard,BPS,Yama,Spain}.
In our approach we do not use the particles exchange operators.}

The relation to KZ equations is modified in the elliptic case as
follows. To the $R$-matrix-valued Lax pair (\ref{a07})-(\ref{a08})
let us associate the ${\rm gl}_{\ti N}$ KZB equations on elliptic
curve with $N$ punctures at $q_i$:
 \beq\label{a75}
 \begin{array}{c}
  \displaystyle{
 \nabla_i\psi=0\,, \ \ \nabla_i=\hbar\,\p_i+\nu\sum\limits_{j:j\neq i}
r_{ij}(q_i-q_j)\,,
 }
 \\
   \displaystyle{
 \nabla_\tau\psi=0\,, \ \
 \nabla_\tau=\hbar\,\p_\tau+\frac{\nu}{2}\sum\limits_{j\neq k}
m_{jk}(q_j-q_k)\,,
 }
 \end{array}
 \eq
where $r_{ij}$ is the classical $r$-matrix, $m_{ij}$ is the next
term in the classical limit (\ref{a1005}), and $i=1,...,N$. The
commutativity of the connections follows from the classical
Yang-Baxter equation
 \beq\label{a760}
 \begin{array}{c}
  \displaystyle{
 [r_{ab},r_{ac}]+[r_{ab},r_{bc}]+[r_{ac},r_{bc}]=0
 }
 \end{array}
 \eq
and the identities
 \beq\label{a76}
 \begin{array}{c}
  \displaystyle{
 [r_{ab},m_{ac}+m_{bc}]+[r_{ac},m_{ab}+m_{bc}]=0\,,
 }
 \end{array}
 \eq
which are obtained from the associative Yang-Baxter equation
(\ref{a16}) and (\ref{a18})-(\ref{a19}). It was shown in \cite{LOZ}
that a solution of (\ref{a75}) satisfies also
  \beq\label{a77}
 \begin{array}{c}
  \displaystyle{
-N\nu\hbar\,\p_\tau\psi=\left(\frac12\sum\limits_{i=1}^N\,\hbar^2\p^2_{q_i}
-\nu^2 {\ti N}^2
\sum\limits_{i<j}\wp(q_i-q_j)+\hbar\,\nu\sum\limits_{i<j}\p_i
r_{ij}(q_i-q_j)+const\right)\psi
 }
 \end{array}
 \eq
or, equivalently (\ref{a1007})
  \beq\label{a78}
 \begin{array}{c}
  \displaystyle{
-N\nu\hbar\,\p_\tau\psi=\left(H(\hat{p},q)+\hbar\,\nu\mF^0+const\right)\psi\,.
 }
 \end{array}
 \eq
Therefore, in contrast to trigonometric case
(\ref{a72})-(\ref{a73}), the conformal block solving the KZ(B)
equations in the elliptic case is not an eigenfunction of the
quantum Hamiltonian, but a solution of quantum non-autonomous
problem, or the non-stationary Shr\"odinger equation with the same
quantum Hamiltonian.

In order to clarify the link between the KZ(B) equations and the
$R$-matrix-valued Lax pairs, a precise relation is needed between
the conformal blocks and the Baker-Akhiezer  functions of the
corresponding Lax pair. This issue remains open.





\section{Spin exchange operators from Hitchin systems}
\setcounter{equation}{0}

The purpose of this Section is to explain the classical origin of
the $\mF^0$ term (related to the elliptic $R$-matrix) in the
Hamiltonian (\ref{a71}) and/or (\ref{a2020}). We start with a brief
description of the classical spin Calogero-Moser model, and then
proceed to the model of $N$ interacting ${\rm SL}_{\ti N}$ tops,
which provides the classical analogue of the Hamiltonian with the
(elliptic) anisotropic spin exchange operator.

\paragraph{Spin Calogero-Moser model.} In classical mechanics the
 ${\rm sl}_N$ spin Calogero-Moser models are described as follows \cite{BAB} (see also \cite{ABB,Bernard}). The
phase space is a direct product $\mC^{2N}\times{\mathcal O}$: the
particles degree of freedom $p_i,q_i$, $i=1...N$ parameterize
$\mC^{2N}$ and the coadjoint orbit of ${\rm GL}_N$ group ${\mathcal
O}$ is parameterized by $S_{ij}$, $i,j=1...N$, i.e. the classical
spin variables are arranged into matrix
$S=\sum_{i,j}E_{ij}S_{ij}\in{\rm Mat}_N$ with fixed eigenvalues. The
Poisson structure is a direct sum of the canonical brackets
(\ref{a02}) and the Lie-Poisson structure
  \beq\label{a80}
  \begin{array}{c}
  \displaystyle{
\{S_{ij},S_{kl}\}=-S_{il}\delta_{kj}+S_{kj}\delta_{il}\,.
 }
 \end{array}
 \eq
The latter is realized by embedding $S$ into a (larger) space with
canonical variables:
  \beq\label{a81}
  \begin{array}{c}
  \displaystyle{
 S_{ij}=\sum\limits_{a=1}^{\ti N}\xi_i^a\eta_j^a\,,\quad\quad
 \{\xi_i^a,\eta_j^b\}=\delta_{ij}\delta_{ab}\,.
 }
 \end{array}
 \eq
The Lax pair
  \beq\label{a82}
  \begin{array}{c}
  \displaystyle{
L^{\hbox{\tiny{spin}}}_{ij}(z)=\delta_{ij}(p_i+S_{ii}E_1(z))+(1-\delta_{ij})S_{ij}\phi(z,q_{ij})\,,
 }
 \end{array}
 \eq
  \beq\label{a86}
  \begin{array}{c}
  \displaystyle{
  M_{ij}^{\hbox{\tiny{spin}}}(z)=(1-\delta_{ij})S_{ij}f(z,q_i-q_j)
 }
 \end{array}
 \eq
and the Hamiltonian
  \beq\label{a87}
  \begin{array}{c}
  \displaystyle{
H^{\hbox{\tiny{spin}}}=\sum\limits_{i=1}^N\frac{p_i^2}{2}-\sum\limits_{i>j}^N
S_{ij}S_{ji}E_2(q_i-q_j)\,.
 }
 \end{array}
 \eq
satisfy the Lax equations (\ref{a06}) and define integrable system
if the additional constraints hold:\footnote{The expressions for the
$M$-matrix (\ref{a86}) is valid only before the Poisson reduction is
performed, which may led to additional (Dirac) terms. For instance,
if $\ti N=1$ in (\ref{a81}) (the minimal orbit) then this reduction
kills all spin variables, and the model is reduced to the spinless
Calogero-Moser model (\ref{a04}). As a result the diagonal terms in
(\ref{a05}) appear.}
%
  \beq\label{a85}
  \begin{array}{c}
  \displaystyle{
 S_{ii}=\hbox{const}\,,\quad \hbox{for all}\ i\,.
 }
 \end{array}
 \eq
The are generated by the action of the Cartan subgroup of ${\rm
SL}_N$.

The appearance of the spin exchange operator (\ref{a67}) comes from
rewriting $S_{ij}S_{ji}$ in the Hamiltonian in terms of $N$ ${\rm
Mat}_{\ti N}$-valued matrices:
  \beq\label{a88}
  \begin{array}{c}
  \displaystyle{
S_{ij}S_{ji}=\sum\limits_{a,b=1}^{\ti
N}\xi_i^a\eta_j^a\xi_j^b\eta_i^b=\tr(\stackrel{i}{B}\stackrel{j}{B})\,,\quad
 \stackrel{i}{B}=\sum\limits_{a,b=1}^{\ti N}{\ti E}_{ab}\stackrel{i}{B}_{ab}\in {\rm Mat}_{\ti
 N}\,,\quad \stackrel{i}{B}_{ab}=\xi_i^a\eta_i^b\,.
 }
 \end{array}
 \eq
Due to (\ref{a85}) $\tr \stackrel{i}{B}=\hbox{const}$ for all
$i=1,...,N$. Hence, we come to $N$ minimal coadjoint orbits of ${\rm
GL}_{\ti N}$. The quantization of $\stackrel{i}{B}_{ab}$ in the
fundamental representation of ${\rm gl}_{\ti N}$ is given by
$\stackrel{i}{\ti E}_{ab}$. This provides the permutation operator
$P_{ij}=\sum\limits_{a,b=1}^{\ti
N}\stackrel{i}{E}_{ab}\stackrel{j}{E}_{ba}$ for quantization of
$\tr(\stackrel{i}{B}\stackrel{j}{B})=\sum\limits_{a,b=1}^{\ti
N}\stackrel{i}{B}_{ab}\stackrel{j}{B}_{ba}$.

Let us remark that the classical $r$-matrix structure for
(\ref{a82}) is defined by dynamical $r$-matrix. Its quantization is
performed in terms of quantum dynamical $R$-matrix \cite{ABB}, while
the approach using the quantum Lax pairs of (\ref{a65})-(\ref{a66})
type is more like the Matsuo-Cherednik construction. It is dual to
the direct one since it is based on ${\rm gl}_{\ti N}$ KZ equations,
and the positions of particles are the $N$ punctures on the base
spectral curve.

\paragraph{The classical version of the $\mF^0$ term}
(\ref{a71}) coming from the elliptic $R$-matrix (\ref{a1004}) is as
follows:
  \beq\label{a811}
  \begin{array}{c}
  \displaystyle{
 \mF^0=\frac12\,\sum\limits_{i\neq j}^NF^0_{ij}(q_{ij})\stackrel{(\ref{a1007})}{=}
 \frac12\,\sum\limits_{i\neq j}^N\p_{q_i}r_{ij}(q_{ij})\,,
 }
  \end{array}
 \eq
 where using
 (\ref{a1006}) and (\ref{a9071}) we have
  \beq\label{a812}
  \begin{array}{c}
  \displaystyle{
 \p_{q_i}r_{ij}(q_{ij})=
 }
 \\ \ \\
  \displaystyle{
  -E_2(q_{ij}) \stackrel{i}{T}_{0}\,\stackrel{j}{T}_{0}
 +\sum\limits_{\ga\in\,\mZ_{\ti N}\times\mZ_{\ti N}\,,\ga\neq0}\vf_\ga(q_{ij},\om_\ga)
 (E_1(q_{ij}+\om_\ga)-E_1(q_{ij})+2\pi\imath\p_\tau\om_\ga)
 \stackrel{i}{T}_{\ga}\,\stackrel{j}{T}_{-\ga}\,.
 }
 \end{array}
 \eq
 It follows from the previous discussion that the classical analogue
 of (\ref{a811})-(\ref{a812}) is
  \beq\label{a813}
  \begin{array}{c}
  \displaystyle{
 \mF^0_{\hbox{\tiny{class}}}=
 }
 \\
  \displaystyle{
 \frac12\,\sum\limits_{i\neq j}^N\Big( -\!E_2(q_{ij})\! \stackrel{i}{B}_{0}\,\stackrel{j}{B}_{0}
 +\!\sum\limits_{\ga\neq0}\vf_\ga(q_{ij},\om_\ga)
 (E_1(q_{ij}+\om_\ga)-E_1(q_{ij})+2\pi\imath\p_\tau\om_\ga)
 \stackrel{i}{B}_{\ga}\,\stackrel{j}{B}_{-\ga}\!\Big)\,,
 }
 \end{array}
 \eq
where $\stackrel{i}{B}$ are the $N$ ${\rm Mat}_{\ti N}$-valued
matrices (classical spin variables), and $\stackrel{i}{B}_\al$ --
their components in the basis (\ref{a904}).

\paragraph{Hitchin systems on ${\rm SL}_{N\ti N}$-bundles.}
In the Hitchin approach \cite{Hi} to elliptic models \cite{Nekr} the
positions of particles are treated as coordinates on the moduli
space of Higgs bundles over (punctured) elliptic curve. The ${\rm
GL}_n$-bundles are classified according to Atiyah \cite{Atiyah}.
Dimensions of the moduli spaces are obtained as the greatest common
divisors of the rank $n$ and degrees of underlying holomorphic
vector bundles. By changing the degrees we describe different type
models. This leads to integrable systems called interacting elliptic
tops \cite{LZ}. Similar construction exists for arbitrary complex
simple Lie group \cite{LOSZ}. The classification is based on
characteristic classes, which are in one to one correspondence with
the center of the structure group.

Here we consider the model related to the ${\rm SL}({N\ti
N},\mC)$-bundle and the characteristic class is $N$ (or
$\exp(2\pi\imath\frac{N}{N\ti N})$). It means that the model
contains $N$ particles degrees of freedom, and the Lax matrix is
${\rm Mat}_{N{\ti N}}$-valued:
  \beq\label{a814}
  \begin{array}{c}
  \displaystyle{
 L(z)=\sum\limits_{i,j=1}^N E_{ij}\otimes L^{ij}(z)\,,\quad\quad
 L^{ij}(z)\in{\rm Mat}_{\ti N}\,,
 }
  \end{array}
 \eq
where
  \beq\label{a815}
  \begin{array}{c}
  \displaystyle{
 L^{ij}(z)=
 }
  \\ \ \\
  \displaystyle{
 =\delta_{ij}\Big(p_i+\mS^{\,ii}_0\, T_0\,E_1(z)+\!\sum\limits_{\al\neq0}\mS^{\,ii}_\al\, T_\al\,\vf_\al(z,\om_\al)\Big)
  +(1-\delta_{ij})\sum\limits_{\al}\mS^{\,ij}_\al\,
  T_\al\,\vf_\al(z,\om_\al-\frac{q_{ij}}{\ti N})\,,
 }
  \end{array}
 \eq
where $T_\al$ is the basis (\ref{a904}) and $\mS^{\,ij}_\al$ -- the
components in this basis of $N^2$ matrices $\mS^{\,ij}\in{\rm
Mat}_{\ti N}$. The variables $\mS^{\,ij}_\al$  are dual to the basis
$E_{ij}\otimes T_\al$ in ${\rm Mat}_{N\ti N}$. Therefore, the
Poisson brackets are defined in the same way as the commutation
relations for the set $\{E_{ij}\otimes T_\al\}$, i.e.
  \beq\label{a8151}
  \begin{array}{c}
  \displaystyle{
 \{\mS^{\,ij}_\al,\mS^{\,kl}_\be\}=\delta_{il}\,\kappa_{\al,\be}\,\mS^{\,kj}_{\al+\be}-
 \delta_{kj}\,\kappa_{\be,\al}\,\mS^{\,il}_{\al+\be}\,.
 }
  \end{array}
 \eq
Of course $\mS^{\,ij}$ can be decomposed in the standard basis in
${\rm Mat}_{\ti N}$ as well. Then the brackets (\ref{a8151}) acquire
the form:
  \beq\label{a8152}
  \begin{array}{c}
  \displaystyle{
 \{\mS^{\,ij}_{ab},\mS^{\,kl}_{cd}\}=\delta_{il}\,\delta_{ad}\,\mS^{\,kj}_{cb}
 -\delta_{kj}\,\delta_{cb}\,\mS^{\,il}_{ad}\,,\quad\quad\mS^{\,ij}=\sum\limits_{a,b,=1}^{\ti N}{\ti
E}_{ab}\,\mS^{\,ij}_{ab}\,.
 }
  \end{array}
 \eq
The analogue for constraints (\ref{a85}) is again generated by the
coadjoint action of the Cartan subgroup of ${\rm GL}_{N}\subset{\rm
GL}_{N{\ti N}}$:
  \beq\label{a816}
  \begin{array}{c}
  \displaystyle{
 \mS^{\,ii}_0=\hbox{const}\quad \hbox{for all}\ i=1,...,N\,.
 }
  \end{array}
 \eq
The quadratic Hamiltonian is evaluated from $\tr L^2(z)$:
  \beq\label{a817}
  \begin{array}{c}
  \displaystyle{
 H^{\rm tops}=\frac{1}{2}\sum\limits_{i=1}^N
 p_i^2-\frac{1}{2}\sum\limits_{i=1}^N\sum\limits_{\al\neq 0}
 \mS^{\,ii}_\al\, \mS ^{ii}_{-\al}\,E_2(\om_\al)-
 \frac{1}{2}\sum\limits_{i\neq j}^N\sum\limits_{\al}
 \mS^{\,ij}_{\al}\,\mS ^{ji}_{-\al}\,E_2(\om_\al-\frac{q_{ij}}{\ti N})\,.
 }
  \end{array}
 \eq
Consider the case of the minimal coadjoint orbit of ${\rm GL}_{N\ti
N}$. Then the matrix
  \beq\label{a818}
  \begin{array}{c}
  \displaystyle{
 \mS=\sum\limits_{i,j=1}^N E_{ij}\otimes\mS^{\,ij}\,,\quad \mS^{\,ij}\in{\rm Mat}_{\ti
 N}\,,\quad i,j=1,...,N
 }
  \end{array}
 \eq
 is of rank 1, that is
  \beq\label{a819}
  \begin{array}{c}
  \displaystyle{
 \mS^{\,ij}=\sum\limits_{a,b,=1}^{\ti N}{\ti E}_{ab}\,\xi^a_i\,\eta^b_j
 }
  \end{array}
 \eq
with $\xi^a_i$ and $\eta^b_j$ are as in (\ref{a81}). This
parametrization is in agreement with the Poisson brackets
(\ref{a8152}) or (\ref{a8151}). From (\ref{a88}) we also have
  \beq\label{a820}
  \begin{array}{c}
  \displaystyle{
 \mS^{\,ii}=\stackrel{i}{B}\,.
 }
  \end{array}
 \eq
Due to (\ref{a9050}) $\mS^{\,ij}_\al=\tr(\mS^{\,ij}\, T_{-\al})/{\ti
N}$. Therefore, using the ${\ti N}$-dimensional columns $\xi^i$ and
the ${\ti N}$-dimensional rows $\eta^j$ we have
  \beq\label{a821}
  \begin{array}{c}
  \displaystyle{
 \mS^{\,ij}_{\al}\,\mS ^{ji}_{-\al}=\frac{\tr(\eta^j\, T_{-\al}\,\xi^i)\,\tr(\eta^i\,
 T_{\al}\,\xi^j)}{{\ti N}^2}=\frac{\tr(\eta^j\, T_{-\al}\,\xi^i\,\eta^i\,
 T_{\al}\,\xi^j)}{{\ti N}^2}=\frac{\tr(\stackrel{i}{B}\,T_{\al}\stackrel{j}{B}T_{-\al})}{{\ti N}^2}\,.
 }
  \end{array}
 \eq
Plugging this expression into (\ref{a817}) we get
  \beq\label{a822}
  \begin{array}{c}
  \displaystyle{
 \!H^{\rm tops}\!=\frac{1}{2}\sum\limits_{i=1}^N
 p_i^2-\frac{1}{2}\sum\limits_{i=1}^N\sum\limits_{\al\neq 0}
 \stackrel{i}{B}_\al\, \stackrel{i}{B}_{-\al}E_2(\om_\al)-
 \frac{1}{2}\sum\limits_{i\neq j}^N\sum\limits_{\al}
 \frac{\tr(\stackrel{i}{B}T_{\al}\!\stackrel{j}{B}\!T_{-\al})}{{\ti N}^2}E_2(\om_\al-\frac{q_{ij}}{\ti N})\,.
 }
  \end{array}
 \eq
The Hamiltonians of this type were originally found in \cite{Polych}
from the study of matrix models. In \cite{LZ} the model (\ref{a822})
was called the interacting elliptic tops, since the second term is
the sum of Hamiltonians for  ${\rm SL}_{\ti N}$ elliptic integrable
tops, and the last term describes their interaction.

Let us write the last term in (\ref{a822}) more explicitly. For this
purpose plug $\stackrel{i}{B}=\sum_\ga T_\ga\!\stackrel{i}{B}_\ga$
and $\stackrel{j}{B}=\sum_\mu T_\mu\!\stackrel{j}{B}_\mu$ into
(\ref{a822}):
  \beq\label{a823}
  \begin{array}{c}
  \displaystyle{
 \sum\limits_{i\neq j}^N\sum\limits_{\al}
 \tr(\stackrel{i}{B}\,T_{\al}\stackrel{j}{B}T_{-\al})E_2(\om_\al-\frac{q_{ij}}{\ti
 N})=\sum\limits_{i\neq
 j}^N\sum\limits_{\al,\mu,\ga}\kappa_{\al,\mu}^2\!
 \stackrel{i}{B}_\ga\stackrel{j}{B}_\mu
 E_2(\om_\al-\frac{q_{ij}}{\ti N})\,\tr(T_\ga T_\mu)=
 }
 \\ \ \\
  \displaystyle{
 \stackrel{(\ref{a9050})}{=}{\ti N}\sum\limits_{i\neq
 j}^N\sum\limits_{\al,\mu}\kappa_{\al,\mu}^2\!
 \stackrel{i}{B}_{-\mu}\stackrel{j}{B}_\mu
 E_2(\om_\al-\frac{q_{ij}}{\ti N})=
 {\ti N}\sum\limits_{i\neq
 j}^N\sum\limits_{\al,\mu}\kappa_{\al,\mu}^2\!
 \stackrel{i}{B}_{\mu}\stackrel{j}{B}_{-\mu}
 E_2(\om_\al+\frac{q_{ij}}{\ti N})\,,
 }
  \end{array}
 \eq
where we used $T_\al T_\mu T_{-\al}=\kappa_{\al,\mu}^2 T_\mu$ coming
from (\ref{a905}), and changed the indices $i,j$ in the last line.
Finally, we sum up over the index $\al$ by applying the Fourier
transformation formulae (\ref{a98})-(\ref{a99}). This yields
  \beq\label{a824}
  \begin{array}{c}
  \displaystyle{
 \frac12\,\sum\limits_{i\neq j}^N\sum\limits_{\al}
 \tr(\stackrel{i}{B}\,T_{\al}\stackrel{j}{B}T_{-\al})E_2(\om_\al-\frac{q_{ij}}{\ti
 N})=-{\ti N}^3 \mF^0_{\hbox{\tiny{class}}}\,,
 }
  \end{array}
 \eq
where $\mF^0_{\hbox{\tiny{class}}}$ was defined in (\ref{a813}). The
answer for the Hamiltonian (\ref{a822}) or (\ref{a817}) is as
follows
  \beq\label{a825}
  \begin{array}{c}
  \displaystyle{
 H^{\rm tops}=\frac{1}{2}\sum\limits_{i=1}^N
 p_i^2-\frac{1}{2}\sum\limits_{i=1}^N\sum\limits_{\al\neq 0}
 \stackrel{i}{B}_\al\, \stackrel{i}{B}_{-\al}E_2(\om_\al)+
 {\ti N}\mF^0_{\hbox{\tiny{class}}}\,.
 }
  \end{array}
 \eq
Hence, we showed that the classical analogue of the anisotropic
elliptic spin exchange operator (see (\ref{a71}), (\ref{a78}),
(\ref{a811})) arises in the model (\ref{a814})-(\ref{a817}) likewise
the classical version of the isotropic spin exchange operator
(\ref{a88}) comes from the spin Calogero model (\ref{a82}),
(\ref{a87}).

The proof of Proposition \ref{2222} comes from quantization of
(\ref{a825}).
First, notice that the second term turns into a constant term
proportional to identity matrix (in ${\rm Mat}_{\ti N}^{\otimes N}$)
when quantized in the fundamental representation since
$\stackrel{i}{T}_\al\, \stackrel{i}{T}_{-\al}=1_{\ti N}$. It could
be added to the original definition of $\mF^0$ (\ref{a09}) as the
terms corresponding to $m=k$ under the sum. The potential of the
Calogero-Moser model comes from the scalar part (the first term in
(\ref{a812})) of the $\mF^0$ term (\ref{a811}). The difference
between $E_2$ and $\wp$-functions (\ref{a908}) is also included into
the constant term proportional to identity matrix. In this way come
to the statement (\ref{a2222}). Similarly, in classical mechanics
the Calogero-Moser potential comes from the first term of
(\ref{a813}) since $\stackrel{i}{B}_{0}$ are equal to the same
constant (\ref{a816}) for all $i$. $\blacksquare$



\section{Conclusion}
\setcounter{equation}{0}


We considered $R$-matrix-valued Lax pairs for $N$-body
Calogero-Moser models. The one for ${\rm A}_{N-1}$ root system was
previously known \cite{LOZ}. We proposed their extensions to other
root systems. Namely, we studied generalizations of the
D'Hoker-Phong Lax pairs \cite{DP} for the classical roots systems in
the untwisted case. These Lax pairs  are block-matrices of $2N\times
2N$ or $(2N+1)\times (2N+1)$ size, and each block is of the size
 ${\ti N}^r\times {\ti N}^r$, where $r$ -- is the number of quantum spaces (spin sites).
Two possibilities were considered. The first one is to keep all $2N$
(or $2N+1$) quantum spaces in $R$-matrices. This leads to the Lax
pairs for ${\rm C}_N$ and ${\rm BC}_N$ cases. The second possibility
is to leave only half ($N$ or $N+1$) quantum spaces. It results in
constructing ${\rm B}_N$ and ${\rm D}_N$ models with ${\rm GL}_2$
($\ti N=2$) Baxter's $R$-matrix. The summary of admissible values of
the coupling constants and the number of quantum spaces in
$R$-matrices is presented in the table below (horizontally are the
numbers of quantum spaces).

 \begin{center}
\begin{tabular}{|c|c|c|c|c|}
 \hline      & N & N+1 & 2N & 2N+1  \\
 \hline & $g=0$, $\mu = 0 $ & & &\\
SO(2N) & $\quad\tilde{N}= 2\quad$    &   &     &      \\
 \hline & & $g = \pm\sqrt{2}\nu$, $\mu = 0 $ & & \\ SO(2N+1)
& & $\tilde{N}= 2$  & &  \\
 \hline & & & $g=0$, $\mu = \nu $ &\\
Sp(2N) &  &   & $\quad\tilde{N}=\hbox{any}\quad$    & \\
 \hline &  & & & $g = \pm \nu$, $\mu = \nu $ \\
BC(N)  &  &    &   &  $\tilde{N}=$  any
\\   \hline
\end{tabular}

{\sf Number of spin quantum spaces and values of coupling
constants.}
 \end{center}

\vskip3mm

Recall that the ordinary Lax pairs (\ref{a31})-(\ref{a322}) were
defined for the following values of the coupling constants:

-- ${\rm SO}_{2N}$: $\mu=0$, $g=0$;

-- ${\rm SO}_{2N+1}$: $\mu=0$, $g^2=2\nu^2$;

-- ${\rm Sp}_{2N}$: $g=0$;

-- ${\rm BC}_{N}$: $g(g^2 - 2\nu^2 + \nu \mu) = 0$.

In this respect our results are as follows: the $R$-matrix-valued
ansatz generalizing (\ref{a31})-(\ref{a322}) works with additional
constraints. For ${\rm SO}$ cases the additional condition is $\ti
N=2$, while for ${\rm C}_{N}$ and ${\rm BC}_{N}$ cases there is no
restriction on $\ti N$ but the constants should satisfy $\mu=\nu$
together with $g=0$ or $g=\pm\nu$ for ${\rm C}_{N}$ or ${\rm
BC}_{N}$ root systems respectively.

Then we proceed to the quantum Lax pairs. A short summary is that
the classical $R$-matrix-valued Lax pairs are generalized to quantum
Lax pairs only for ${\rm SO}$ cases from the above table.

The quantum Lax pairs are naturally related to the spin
Calogero-Moser models. The corresponding spin exchange operators
$\mF^0$ appear as a scalar parts of the $R$-matrix-valued
$M$-matrices. On the other hand the same operators can be derived
from KZ or KZB equations. We demonstrate these relations for ${\rm
sl}_N$ $R$-matrix-valued Lax pair. The link between the
operator-valued Lax pairs and KZ equations comes from the
Matsuo-Cherednik duality. Its quasi-classical version provides the
so-called quantum-classical duality between the quantum spin chains
(Gaudin models) and the classical many-body systems of
Ruijsenaars-Schneider (Calogero-Moser) type \cite{GZZ}. In this
paper we deal with another example of quantum-classical relation.
We treat the Lax equations for the classical Calogero-Moser model
(\ref{a07})-(\ref{a09}) with $R$-matrix-valued Lax pairs as
half-quantum model (\ref{a2020}), which quantum part is described by
the spin exchange operator known previously as the "noncommutative
spin interactions" \cite{Polych}. The spin variables are quantized
in the fundamental representation, while the particles degrees of
freedom remain classical. We show that the classical counterpart of
the elliptic anisotropic spin exchange operator comes from the
Hitchin type system on ${\rm SL}_{N\ti N}$-bundle with nontrivial
characteristic class over elliptic curve. See the Proposition
\ref{2222}.

It was shown in \cite{SeZ} that the spin exchange operator $\mF^0$
(\ref{a812}) for $\ti N=2$ being reduced to the equilibrium position
$q_j=j/N$ provides the Hamiltonian for anisotropic extension of the
Inozemtsev elliptic long-range chain.
In view of the relation of the $R$-matrix-valued Lax pairs and the
Hitchin systems on ${\rm SL}(N\ti N)$-bundles we expect that these
type long-range integrable spin chains admit Lax representations of
size $N\ti N\times N\ti N$ at both - classical and quantum levels.
They are obtained from the one for interacting tops (\ref{a815}) by
the substitution $p_j=0$, $q_j=j/N$. Such Lax pair allows to
calculate the higher Hamiltonians. These questions will be discussed
in our next publication \cite{GSZ}.





\section{Appendix}

\def\theequation{A.\arabic{equation}}
\setcounter{equation}{0}

In addition to the standard basis in ${\rm Mat}_{\ti N}$ we use the
one \cite{Belavin}
 \beq\label{a904}
 \begin{array}{c}
  \displaystyle{
 T_a=T_{a_1 a_2}=\exp\left(\frac{\pi\imath}{{\ti N}}\,a_1
 a_2\right)Q^{a_1}\Lambda^{a_2}\,,\quad
 a=(a_1,a_2)\in\mZ_{\ti N}\times\mZ_{\ti N}
 }
 \end{array}
 \eq
constructed by means of the finite dimensional representation of
Heisenberg group
 \beq\label{a903}
 \begin{array}{c}
  \displaystyle{
Q_{kl}=\delta_{kl}\exp(\frac{2\pi
 \imath}{{\ti N}}k)\,,\ \ \ \Lambda_{kl}=\delta_{k-l+1=0\,{\hbox{\tiny{mod}}}
 {\ti N}}\,,\quad Q^{\ti N}=\Lambda^{\ti N}=1_{{\ti N}\times {\ti N}}\,.
 }
 \end{array}
 \eq
 The following relations hold
  \beq\label{a905}
 \begin{array}{c}
  \displaystyle{
T_\al T_\be=\kappa_{\al,\be} T_{\al+\be}\,,\ \ \
\kappa_{\al,\be}=\exp\left(\frac{\pi \imath}{{\ti N}}(\be_1
\al_2-\be_2\al_1)\right)\,,
 }
 \end{array}
 \eq
  \beq\label{a9050}
 \begin{array}{c}
  \displaystyle{
\tr(T_\al T_\be)={\ti N}\delta_{\al,-\be}\,,
 }
 \end{array}
 \eq
where $\al+\be=(\al_1+\be_1,\al_2+\be_2)$. The permutation operator
takes the form
  \beq\label{a9051}
 \begin{array}{c}
  \displaystyle{
P_{12}=\sum\limits_{i,j=1}^{\ti N} {\ti E}_{ij}\otimes {\ti
E}_{ji}=\frac{1}{{\ti N}}\sum\limits_{\al\in\,\mZ_{\ti
N}\times\mZ_{\ti N}} T_\al \otimes T_{-\al}\,,
 }
 \end{array}
 \eq
where ${\ti E}_{ij}$ is the standard basis in ${\rm Mat}_{\ti N}$.

\vskip4mm

The Kronecker function is defined in the rational, trigonometric
(hyperbolic) and elliptic case as follows:
  \beq\label{a907}
  \begin{array}{l}
  \displaystyle{
 \phi(\eta,z)=\left\{
   \begin{array}{l}
    1/\eta+1/z\quad - \quad \hbox{rational case}\,,
    \\
    \coth(\eta)+\coth(z) \quad - \quad \hbox{trigonometric case}\,,
    \\
    \frac{\vth'(0)\vth(\eta+z)}{\vth(\eta)\vth(z)} \quad - \quad
    \hbox{elliptic
    case}\,.
   \end{array}
 \right.
 }
 \end{array}
 \eq
 In the latter case the theta-function is the odd one
   \beq\label{a9072}
  \begin{array}{c}
  \displaystyle{
   \vth(z)=\displaystyle{\sum _{k\in \mathbb Z}} \exp \left ( \pi
\imath \tau (k+\frac{1}{2})^2 +2\pi \imath
(z+\frac{1}{2})(k+\frac{1}{2})\right )\,.
 }
 \end{array}
 \eq
Similarly, the first Eisenstein (odd) function and the Weierstrass
(even) $\wp$-function:
  \beq\label{a908}
  \begin{array}{c}
  \displaystyle{
 E_1(z)=\left\{
   \begin{array}{l}
 1/z\,,
\\
   \coth(z)\,,
\\
    \vth'(z)/\vth(z)\,,
   \end{array}
 \right.\hskip12mm  \wp(z)=\left\{
   \begin{array}{l}
 1/z^2\,,
\\
   1/\sinh^2(z)\,,
\\
    -\p_z E_1(z)+\frac{1}{3}\frac{\vth'''(0)}{\vth'(0)}\,.
   \end{array}
 \right.
 }
 \end{array}
 \eq
The derivative
   \beq\label{a9081}
  \begin{array}{c}
  \displaystyle{
   E_2(z)= -\p_z E_1(z)
 }
 \end{array}
 \eq
 is the second Eisenstein function.
The derivative of the Kronecker function:
  \beq\label{a9071}
  \begin{array}{l}
  \displaystyle{
 f(z,q)\equiv\p_q\phi(z,q)=\phi(z,q)(E_1(z+q)-E_1(q))\,.
 }
 \end{array}
 \eq
Due to the following behavior of $\phi(z,q)$ near $z=0$
  \beq\label{a9077}
  \begin{array}{l}
  \displaystyle{
\phi(z,q)=z^{-1}+E_1(q)+z\,(E_1^2(q)-\wp(q))/2+O(z^2)\,.
 }
 \end{array}
 \eq
 we also have
  \beq\label{a9078}
  \begin{array}{l}
  \displaystyle{
 f(0,q)=-E_2(q)\,.
 }
 \end{array}
 \eq
The Fay trisecant identity:
  \beq\label{a909}
  \begin{array}{c}
  \displaystyle{
\phi(z,q)\phi(w,u)=\phi(z-w,q)\phi(w,q+u)+\phi(w-z,u)\phi(z,q+u).
 }
 \end{array}
 \eq
 For the Lax equations the following degenerations of (\ref{a909}) are needed
  \beq\label{a966}
  \begin{array}{c}
  \displaystyle{
 \phi(z,x)f(z,y)-\phi(z,y)f(z,x)=\phi(z,x+y)(\wp(x)-\wp(y))\,,
 }
 \end{array}
 \eq
  \beq\label{a912}
  \begin{array}{c}
  \displaystyle{
 \phi(\eta,z)\phi(\eta,-z)=\wp(\eta)-\wp(z)=E_2(\eta)-E_2(z)\,.
 }
 \end{array}
 \eq
 Also
  \beq\label{a911}
  \begin{array}{c}
  \displaystyle{
 \phi(z,q)\phi(w,q)=\phi(z+w,q)(E_1(z)+E_1(w)+E_1(q)-E_1(z+w+q))=
 }
 \\ \ \\
  \displaystyle{
 =\phi(z+w,q)(E_1(z)+E_1(w))-f(z+w,q)\,.
 }
 \end{array}
 \eq
The set of ${\ti N}^2$ functions
 \beq\label{a910}
 \begin{array}{c}
  \displaystyle{
 \vf_a^\eta(z)=\exp(2\pi\imath\frac{a_2}{{\ti N}}\,z)\,\phi(z,\eta+\frac{a_1+a_2\tau}{{\ti N}})\,,\quad
 a=(a_1,a_2)\in \mZ_{\ti N}\times\mZ_{\ti N}
 }
 \end{array}
 \eq
is used in the definition of the Baxter-Belavin's
\cite{Baxter,Belavin} elliptic $R$-matrix
 \beq\label{a1004}
 \begin{array}{c}
  \displaystyle{
R_{12}^\eta(z)=\sum\limits_\al T_\al\otimes T_{-\al}
\vf_\al(z,\om_\al+\eta)\,.
  }
 \end{array}
 \eq
The classical limit (behavior near $\eta=0$)
 \beq\label{a1005}
 \begin{array}{c}
  \displaystyle{
R_{12}^\eta(z)=\frac{1\otimes 1}{\eta}+r_{12}(z)+\eta\, m_{12}(z)
+O(\eta^2)
  }
 \end{array}
 \eq
is similar to (\ref{a9077}). The classical $r$-matrix
 \beq\label{a1006}
 \begin{array}{c}
  \displaystyle{
r_{12}(z)=1\otimes 1 E_1(z)+\sum\limits_{\al\neq 0} T_\al\otimes
T_{-\al} \vf_\al(z,\om_\al)
  }
 \end{array}
 \eq
is skew-symmetric due to (\ref{a19}) or (\ref{a18}). From
(\ref{a1005}) we conclude that
 \beq\label{a1007}
 \begin{array}{c}
  \displaystyle{
F^0_{12}(q)=\p_q R^\eta_{12}(q)|_{\eta=0}=\p_q
r_{12}(q)=F^0_{21}(-q)\,.
  }
 \end{array}
 \eq

 The finite Fourier transformation for the set of
functions (\ref{a910}) is as follows (see e.g. \cite{Fourier}):
 \beq\label{a913}
 \begin{array}{c}
  \displaystyle{
\frac{1}{{\ti N}}\sum\limits_{\al} \kappa_{\al,\ga}^2\, \vf_\al({\ti
N}\eta,\om_\al+\frac{z}{{\ti N}})=\vf_\ga(z,\om_\ga+\eta)\,,\quad
\forall\ga\,.
  }
 \end{array}
 \eq
It is generated by the arguments symmetry (similarly to
$\phi(z,q)=\phi(q,z)$)
 \beq\label{a1008}
 \begin{array}{c}
  \displaystyle{
R_{12}^z(q)P_{12}=R_{12}^{q/{\ti N}}({\ti N}z)\,.
  }
 \end{array}
 \eq
 In particular, (\ref{a913}) leads to
  \beq\label{a98}
 \begin{array}{c}
  \displaystyle{
 \sum\limits_{\al} E_2(\om_\al+\eta)={\ti N}^2E_2({\ti N}\eta)
 \quad\hbox{or}\quad
 \sum\limits_{\al} \wp(\om_\al+\eta)={\ti N}^2\wp({\ti N}\eta)
  }
 \end{array}
 \eq
 and for $\ga\neq 0$
  \beq\label{a99}
 \begin{array}{c}
  \displaystyle{
 \sum\limits_{\al} \kappa_{\al,\ga}^2 E_2(\om_\al+\eta)=-
 {\ti N}^2\vf_\ga({\ti N}\eta,\om_\ga)
 (E_1({\ti N}\eta+\om_\ga)-E_1({\ti N}\eta)+2\pi\imath\p_\tau\om_\ga)\,.
  }
 \end{array}
 \eq

\vskip2mm

\noindent {\bf Acknowledgments.} We are grateful to A. Levin, M.
Olshanetsky and I. Sechin for helpful discussions. The work was
performed at the Steklov Mathematical Institute of Russian Academy
of Sciences, Moscow. This work is supported by the Russian Science
Foundation under grant 14-50-00005.

\begin{small}

\end{small}

\end{document}